\documentclass[
  journal=pasa,
  manuscript=research-paper, %% or "review"
  year=2026,
  volume=10151,
]{cup-journal}

\doi{10.1017/pasa.2026.10151}

\usepackage{amsmath}
\usepackage[table]{xcolor}
\usepackage{multirow}
\RequirePackage{aas-macros}
\usepackage{orcidlink}
\usepackage{savesym}
\savesymbol{vert}
\usepackage{yhmath}
\restoresymbol{SYM}{vert}
\usepackage{mathtools}
\usepackage[nopatch]{microtype}
\usepackage{booktabs}
\usepackage{listings}
\usepackage{graphicx}
\usepackage{graphbox} 
\usepackage{natbib}
\usepackage{hyperref}
\DeclareUnicodeCharacter{2212}{-}
\usepackage{lineno}
\usepackage{booktabs}
\usepackage{tabularx}
\usepackage{array}
\usepackage{subcaption}
\usepackage{xcolor}

\newcolumntype{C}{>{\centering\arraybackslash}X}
\defcitealias{VanKempen2024}{Paper~I}

\title{Group Therapy for Halos: Advancing Halo Mass Estimation for Galaxy Groups}

\author{Wesley Van Kempen \orcidlink{0009-0009-8499-1326}}
\affiliation{Centre for Astrophysics and Supercomputing, Swinburne University of Technology, John Street, Hawthorn, VIC 3122, Australia}
\email[Wesley Van Kempen]{wvankempen@swin.edu.au}

\author{Michelle E. Cluver \orcidlink{0000-0002-9871-6490}}
\affiliation{Centre for Astrophysics and Supercomputing, Swinburne University of Technology, John Street, Hawthorn, VIC 3122, Australia}
\alsoaffiliation{Department of Physics and Astronomy, University of the Western Cape, Robert Sobukwe Road, Bellville, 7535, South Africa}

\author{Edward N. Taylor \orcidlink{0000-0002-5522-9107}}
\affiliation{Centre for Astrophysics and Supercomputing, Swinburne University of Technology, John Street, Hawthorn, VIC 3122, Australia}

\author{Darren J. Croton \orcidlink{0000-0002-5009-512X}}
\affiliation{Centre for Astrophysics and Supercomputing, Swinburne University of Technology, John Street, Hawthorn, VIC 3122, Australia}
\alsoaffiliation{ARC Centre of Excellence for All Sky Astrophysics in 3 Dimensions (ASTRO 3D), Melbourne, Australia}
\alsoaffiliation{ARC Centre of Excellence for Dark Matter Particle Physics (CDM), Melbourne, Australia}

\author{Trystan S. Lambert \orcidlink{0000-0001-6263-0970}}
\affiliation{International Centre for Radio Astronomy Research (ICRAR), M468, University of Western Australia, 35 Stirling Hwy, Crawley, WA 6009, Australia}

\author{Claudia del P. Lagos \orcidlink{0000-0003-3021-8564}}
\affiliation{International Centre for Radio Astronomy Research (ICRAR), M468, University of Western Australia, 35 Stirling Hwy, Crawley, WA 6009, Australia}
\alsoaffiliation{ARC Centre of Excellence for All Sky Astrophysics in 3 Dimensions (ASTRO 3D), Melbourne, Australia}
\alsoaffiliation{Cosmic Dawn Center (DAWN), Denmark.}

\received {22/08/2025}
\revised  {19/10/2025}
\revised  {08/01/2026}
\accepted {12/01/2026}
\published{22/01/2026}


\begin{document}

\begin{abstract}
Accurate estimation of dark matter halo masses for galaxy groups is central to studies of galaxy evolution and for leveraging group catalogues as cosmological probes. In this work, we present a comprehensive evaluation and calibration of two complementary halo mass estimators: a dynamical estimator based on a modified virial theorem (MVT), and an empirical summed stellar mass to halo mass relation (sSHMR) which uses the summed mass of the three most massive group galaxies as a proxy for halo mass. Using a suite of state-of-the-art semi-analytic models (SAMs; S\textsc{hark}, SAGE, and GAEA) to produce observationally motivated mock light-cone catalogues, we rigorously quantify the accuracy, uncertainty, and model dependence of each method. The MVT halo mass estimator achieves negligible systematic bias (mean $\Delta = -0.01$ dex) and low scatter (mean $\sigma = 0.20$ dex) as a function of the predicted halo mass, with no sensitivity to the SAM baryonic physics. The calibrated sSHMR yields the highest precision, with mean $\Delta = 0.02$ dex and mean $\sigma = 0.14$ dex as a function of the predicted halo mass but exhibits greater model dependence due to its sensitivity to varying baryonic physics and physical prescriptions across the SAMs. We demonstrate the application of these estimators to observational group catalogues, including the construction of the empirical halo mass function and the mapping of quenched fractions in the stellar mass–halo mass plane. We provide clear guidance on the optimal application of each method: the MVT is recommended for GAMA-like surveys ($i < 19.2$) calibrated to $z < 0.1$ and should be used for studies that require minimal model dependence, while the sSHMR is optimal for high-precision halo mass estimation across diverse catalogues with magnitude limits of $Z < 21.2$ or brighter and to redshifts  of $z \leq 0.3$. These calibrated estimators will be of particular value for upcoming wide-area spectroscopic surveys, enabling robust and precise analyses between the galaxy–halo connection and the underlying dark matter distribution.
\end{abstract}

%\noindent 

\section{Introduction}
\label{sec:Introduction}

The dark-matter halo-mass represents one of the most fundamental properties governing galaxy evolution within the hierarchical structure formation paradigm; as galaxies form and evolve within these gravitationally bound dark matter structures, the halo mass strongly influences numerous galaxy properties including star formation histories, morphologies, and chemical enrichment pathways \citep{White1978, Blumenthal1984, Bower2006}. Accurate determination of halo masses therefore underpins our ability to connect theoretical models of structure formation with observational constraints on galaxy evolution across cosmic time.

Given the importance of halo mass in the field of galaxy evolution, accurate measures of halo mass are critical; however, it remains challenging to measure directly for individual groups. Weak gravitational lensing provides accurate and widely adopted halo mass estimates, and is routinely used to validate indirect estimators \citep{Leauthaud2012, Hoekstra2013, Velander2014, Mandelbaum2016, Mandelbaum2018}. Nonetheless, robust weak-lensing halo masses for individual galaxy groups are not universally available at typical survey depths, as the per-object signal-to-noise at group-scale masses is often insufficient without stacking or targeted deep imaging \citep{Leauthaud2012, Velander2014, Viola2015, Simet2017}. Other indirect probes include X-ray emission from hot gas \citep[e.g.,][]{Arnaud2010} and galaxy dynamics \citep[e.g.,][]{Old2014}; these can be applied to individual groups, but carry method-specific systematics that are particularly pronounced in low membership groups. This regime is crucial for tracing the growth of structure within group environments and for understanding the transition from group to cluster environments \citep{Yang2007, Robotham2011}. Current approaches to estimating halo masses face several challenges: weak lensing signals become increasingly difficult to detect for lower-mass systems \citep{Viola2015}, while X-ray observations require the presence of hot gas in hydrostatic equilibrium; a condition typically not met in group-scale environments \citep{Lovisari2021}. Traditional dynamical mass estimators based on the virial theorem assume that systems are both virialised and well-sampled; these assumptions often break down for groups with low multiplicity, particularly when survey completeness is limited \citep{Robotham2011, Old2018, Wojtak2018}. Collectively, these limitations introduce substantial scatter and systematic uncertainty in halo mass estimates, particularly in the low mass/multiplicity group regime where accurate halo mass measurements are most critical for understanding the impact of baryonic feedback on galaxy evolution \citep{Wechsler2018}.

Improved halo mass estimations are required to address several fundamental questions in modern astrophysics. First, the precise shape and amplitude of the halo mass function provides important constraints on cosmological parameters, particularly $\sigma_{8}$ and $\Omega_{m}$ \citep{Tinker2008, Castro2021, Driver2022}. Second, understanding the coevolution of galaxies and their host haloes requires accurate mapping between observable galaxy properties and underlying halo masses \citep{Behroozi2019, Moster2020}. The scatter in the stellar halo mass relation (SHMR) offers insights into the complex baryonic processes driving galaxy assembly, including feedback from supernovae, active galactic nuclei (AGN), and star-formation efficiency \citep{Pillepich2018, Davies2019, Oyarzun2024, Wang2025}. More broadly, measuring the baryonic properties of galaxies as a function of halo mass remains a central goal in galaxy evolution and cosmology. For example, \citet{Chauhan2020} demonstrated that the signatures of AGN feedback on the HI content of haloes are remarkably strong, with variations between simulations exceeding 1 dex. However, subsequent work by \citet{Chauhan2021} showed that the uncertainties inherent in halo mass estimation, particularly when using the HI stacking techniques commonly adopted in the literature, can completely obscure these feedback signatures, makes it impossible to distinguish between different simulation models. These studies further highlight that while dynamical mass estimators perform well for high-multiplicity systems, but are unable to resolve the underlying distribution for low mass, low multiplicity groups. This underscores the critical importance of minimising halo mass uncertainties in order to robustly connect baryonic content and feedback processes to the underlying dark matter halo population.

The advent of next-generation wide-field spectroscopic surveys such as Wide Area Vista Extragalactic Survey \citep[WAVES;][]{Driver2019}, the Dark Energy Spectroscopic Instrument \citep[Desi;][]{DESI2016}, and the 4MOST Hemisphere Survey \citep{ENTaylor23} promises to generate comprehensive catalogues of galaxy groups across cosmic time. These surveys will deliver spectroscopic redshifts for millions of galaxies, where high completeness will facilitate the robust identification of gravitationally bound structures at the group scale where environmental effects demonstrably alter galaxy properties and evolutionary trajectories \citep{Peng2010, Wetzel2013, Davies2019}. However, the scientific potential of these datasets precariously depends on our capacity to accurately translate observable group properties into reliable halo mass estimates \citep{Kravtsov2018, Eckert2020, Tinker2021}. This connection between observable baryonic tracers and the underlying dark matter distribution remains fundamental for quantitatively testing hierarchical structure formation models.

The continuous flow of gas into, within, and out of galaxies is referred to as the baryon cycle and represents a key process governing galaxy evolution \citep{Davé2012, Lilly2013}. Complementary multi-wavelength facilities such as Euclid \citep{Laureijs2011}, the Vera Rubin Observatory \citep{LSST2009}, the Square Kilometre Array \citep[SKA;][]{Dewdney2009}, the Atacama Large Millimeter/submillimeter Array \citep[ALMA;][]{Wootten2009}, the extended Roentgen Survey with an Imaging Telescope Array \citep[eROSITA;][]{Merloni2012}, the Australian Square Kilometre Array Pathfinder \citep[ASKAP;][]{Johnston2008}, and the Very Large Telescope/Multi Unit Spectroscopic Explorer \citep[VLT/MUSE;][]{Bacon2010} capture distinct components of this cycle: stellar content through optical and near-infrared observations, molecular gas reservoirs via millimetre observations, hot gas through X-ray measurements, neutral hydrogen via radio observations, and spatially resolved gas kinematics through integral-field spectroscopy \citep{Saintonge2017, Peroux2020, Tacconi2020}. Pairing these multi-wavelength datasets with spectroscopic information and group catalogues allows us to directly probe the influence of dark matter on the baryon cycle and empirically test how halo properties regulate key processes such as gas accretion rates, star formation efficiency, and feedback-driven outflows across diverse environments and cosmic epochs \citep{Tumlinson2017, Mitchell2020, van_de_Voort2021}. A comprehensive understanding of these complex relationships necessitates both the statistical power of large-scale spectroscopic group catalogues and the detailed multi-wavelength characterisation of baryon cycle components. Nevertheless, precise halo mass measurements remain the critical prerequisite, particularly at the group scale ($10^{12}$--$10^{14}$ M$_{\odot}$) where the interplay between dark matter and baryonic processes most significantly influences galaxy evolution \citep{Behroozi2019, Davies2019, Wang2025}.

Semi-analytical models (SAMs) offer a powerful framework for developing and calibrating such techniques. These models implement physically motivated prescriptions for galaxy formation within dark matter halo merger trees from N-body simulations \citep{Prada2012, Somerville2015, Croton2016, Lagos2018, DeLucia2024}. By producing realistic galaxy populations with known halo properties, SAMs allow us to assess the performance of different mass estimation techniques and quantify their associated uncertainties. Recent advances in SAMs, including improved treatments of gas cooling, star formation, and feedback processes, which has significantly enhanced their ability to reproduce observed galaxy properties across a wide range of environments \citep{Klypin2016, Croton2016, Lacey2016, Lagos2018, Stevens2018, Henriques2020, Lagos2024, DeLucia2024}.

In this paper, we present two complementary approaches to improve observational halo mass estimates for galaxy groups. The first approach addresses limitations in traditional dynamical mass estimators via the virial theorem when applied to low-multiplicity groups by developing a corrective framework that accounts for systematic biases in groups with small velocity dispersions and projected radius measurements. The second method uses the correlation between baryonic mass and dark matter mass, probing the mass relationship between the three most massive galaxies in the halo and that of the halo mass. 

This paper is organised as follows. In Sections~\ref{sec:SGP} \&  \ref{sec:SAMS}, we introduce the observational dataset in which we showcase applications of the halo mass relations and describe the SAMs used to calibrate/validate our relations. Section~\ref{sec:Methods} introduces our two calibrated halo mass estimators; Section~\ref{subsec:Meth1} presents the first approach utilising a modified virial theorem (MVT) to estimate the halo mass, and in Section~\ref{subsec:Meth2} we present a summed stellar-halo mass relation (sSHMR) using the three most massive galaxies in a group as a proxy for halo mass. In Section \ref{sec:ObsApp}, we apply the halo mass estimations to the observational data and demonstrate their performance and use cases. Finally, Section \ref{sec:conclusion} summarises our findings and outlines future directions for halo mass estimation in upcoming large-scale surveys.

Throughout this paper, we adopt a flat $\Lambda$CDM cosmology with parameters: H$_{0}$ = 70 km s$^{-1}$ Mpc$^{-1}$, $\Omega_{M}$ = 0.3, and $\Omega_{\Lambda}$ = 0.7, unless stated otherwise. All halo masses are defined as $M_{200}$, the mass enclosed within a radius where the mean density is 200 times the critical density of the Universe.

\section{Observational Data}
\label{sec:SGP}

The primary observational dataset used in this work is the Southern Galactic Pole (SGP) catalogue introduced in \citet{VanKempen2024}. The SGP provides a highly complete spectroscopic sample of galaxies at $z < 0.1$ across 376~deg$^2$ ($340^\circ < \mathrm{RA} < 26^\circ$, $-35.3^\circ < \mathrm{Dec} < -25.8^\circ$), the Two-degree-Field Galaxy Redshift Survey \citep[2dFGRS;][]{Colless01} and Galaxy And Mass Assembly \citep[GAMA;][]{Driver2009} G23 survey regions. Redshifts are sourced primarily from 2dFGRS and GAMA, and are supplemented with the Six-degree Field Galaxy Survey \citep[6dFGRS;][]{Blake16}, the 2-degree Field Lensing Survey \citep[2dFLenS;][]{Jones04}, the 2MASS Redshift Survey \citep[2MRS;][]{Macri19}, and the Million Quasars catalogue \citep[MILLIQUAS;][]{Flesch21} measurements. The SGP catalogue is comprised of 24,656 unique spectroscopic sources. These sources were cross-matched with photometry from the Wide-field Infrared Survey Explorer \citep[WISE;][]{Wright2010}, combining both the WISE Extended Source Catalogue \citep[WXSC;][]{Jarrett13, Jarrett2019} and the point-source AllWISE catalogue \citep{Cluver2014, Cluver2020}. The matching between WISE photometry and that of the spectroscopic sources was approximately 93\%, yielding mid-infrared measurements for 22,933 galaxies. The cross-matched WISE photometry enables robust estimates of stellar masses, derived from the W1-based relations of \cite{Jarrett2023}, as well as star formation rates, calculated from W3 and W4 luminosities following the calibrations of \cite{Cluver2025}.

\begin{figure*}[!hbt]
\centering
\includegraphics[width=\linewidth]{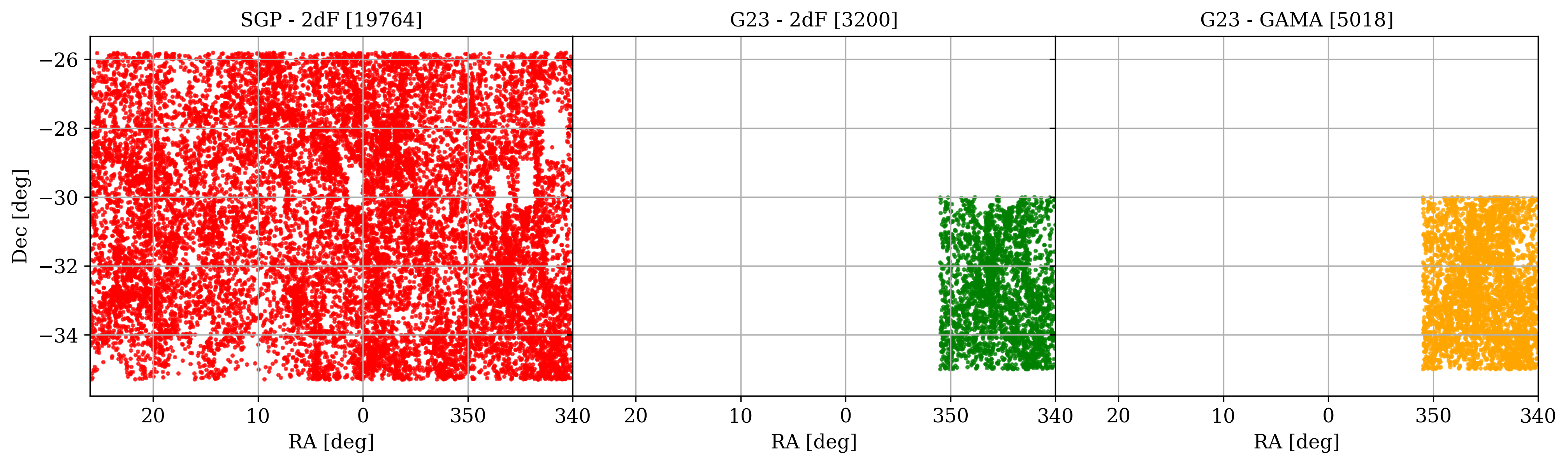}
\caption{RA–Dec distribution of the three SGP sub-samples used for uniform completeness: SGP–2dF (red), G23–2dF (green), and G23–GAMA (orange). Comparing G23–2dF with G23–GAMA highlights the effect of spectroscopic completeness. Numbers in brackets denote sample sizes.}
\label{fig:SGP_Samples}
\end{figure*}

As the SGP catalogue is constructed from multiple spectroscopic catalogues, the resulting dataset is heterogeneous in nature. To establish homogeneity for our analyses, we define three spectroscopic sub-samples: (1) SGP–2dF, comprising all galaxies with 2dFGRS photometry across the full SGP footprint; (2) G23–2dF, containing galaxies with 2dFGRS photometry restricted to the GAMA G23 region ($339^\circ < \mathrm{RA} < 351^\circ$, $-35^\circ < \mathrm{Dec} < -30^\circ$); and (3) G23–GAMA, consisting of galaxies with GAMA photometry within G23. Figure~\ref{fig:SGP_Samples} presents an Right Ascension–Declination view of the homogeneous, WISE cross-matched sub-samples of the SGP dataset. The G23-GAMA sub-samples clearly demonstrate higher spectroscopic completeness, which is approximately 2.3 times greater than that of G23–2dF.

The SGP catalogue provides 1,413 galaxy groups, identified via a Python-based, graph-theory implementation of a friends-of-friends (FoF) algorithm, {\tt FoFpy} \citep{Lambert2020}. {\tt FoFpy} links galaxies when both their projected separations and line-of-sight velocity differences fall below scalable linking lengths, with a probabilistic cut used to reject links (see \citet{Lambert2020} for further details on the {\tt FoFpy} algorithm). To ensure robust recovery of galaxy groups, a two-pass strategy was adopted: an initial FoF run with extended linking lengths captured larger groups, which were removed prior to a second pass with smaller linking lengths targeting smaller groups. The linking lengths were calibrated using mock lightcones constructed from the Millennium simulation \citep{Springel05} and the Semi-Analytic Galaxy Evolution (SAGE) model \citep{Croton2016}. These mock lightcones were generated with 2dFGRS ($b_{J}=19.45$) and GAMA ($i=19.2$) magnitude limits. See \cite{VanKempen2024} for a full description of the construction of the SGP group catalogue.

\section{Semi-Analytic Models of Galaxy Formation}
\label{sec:SAMS}

This section provides a comprehensive overview of the simulated datasets used in this study. We describe the suite of state-of-the-art SAMs of galaxy formation employed in this work. SAMs represent powerful theoretical tools for exploring the interplay between dark matter structure formation and baryonic physics. These models implement physically motivated prescriptions for key astrophysical processes within merger trees extracted from cosmological $N$-body simulations \citep[see reviews by][]{Baugh2006,Benson2010}. By providing self-consistent galaxy populations with fully known dark matter halo properties, SAMs offer an ideal framework for developing and calibrating halo-mass estimation techniques. In this work, three independent SAMs are used with distinct roles: S\textsc{hark} \citep{Lagos2018, Lagos2024} serves as the fiducial calibration model for the halo mass estimates, whereas SAGE and the \textsc{GAlaxy Evolution and Assembly} \citep[GAEA;][]{DeLucia2014, Hirschmann2016, DeLucia2024} models are used exclusively for validation and robustness testing. No parameters of the estimators are re-tuned on SAGE or GAEA. Each of these SAMs implement different physical prescriptions while operating on distinct $N$-body simulations. This multi-model approach enables assessment of robustness to variations in the underlying galaxy-formation physics and cosmology.

To ensure comparability with the observations described in Section~\ref{sec:SGP}, all SAM outputs are post-processed into mock light cones with realistic sky coordinates and redshifts, including peculiar-velocity–induced redshift-space distortions. Apparent magnitudes are computed and an SGP-like selection is applied (e.g., GAMA G23 magnitude limit of $i<19.2$) to emulate the survey depth. Galaxy groups in the SAMs provides ground-truth memberships and halo masses while retaining observational selection effects. Estimator inputs are restricted to observables after the SGP-like selection, this methodology ensures that the calibration and validation of halo mass estimators are performed under conditions that closely mimic real, highly complete spectroscopic surveys, thereby enhancing the reliability and applicability of the methods to current and future observational datasets. As a caveat, a fainter magnitude limit of $Z<21.2$ is used for S\textsc{hark} in the development of the sSHMR (see Section~\ref{subsubsec:SHMR_Acc_Unc}).

\subsection{SHARK}
\label{subsec:SHARK}

S\textsc{hark} v2.0 \citep{Lagos2018, Lagos2024} is an open-source, modular semi-analytic model of galaxy formation and evolution. The latest version incorporates significant advancements, including an improved treatment of angular momentum evolution, several environmental processes including ram pressure and tidal stripping, and updated feedback models. The free parameters in S\textsc{hark} are calibrated to reproduce the observed stellar mass function, star formation rate density, and cold gas scaling relations at $z = 0$.

The S\textsc{hark} runs analysed in this work are based on the SURFS suite of $N$-body simulations \citep{Elahi2018}, specifically medi-SURFS, which spans $(210~h^{-1}~\mathrm{Mpc})^3$ with $1536^3$ dark matter particles, yielding a particle mass of $2.21\times10^8h^{-1}M_{\odot}$. Haloes, sub-haloes, and merger trees are constructed using HBT+HERONS \citep{ChandroGomez2025}. The S\textsc{hark} simulated light cones used in this work correspond to WAVES WIDE (North + South) light cones, totalling $\sim 1,100 \, \mathrm{deg}^2$ in area. These light cones were constructed using the pipeline described in \citet{Lagos2019}: the survey geometry and magnitude selections are built using Stingray \citep{Chauhan2019}, and the galaxy SEDs are built using ProSpect \citep{Robotham2020}. A lower stellar mass limit of $\log M_{\star} > 7.5~[M_{\odot}]$ was applied to the mock light cones to ensure completeness and reliability in the resulting galaxy sample. The simulation adopts a \textit{Planck} 2015 cosmology \citep{Planck2016}, with $\Omega_{\mathrm{m}} = 0.3121$, $\Omega_{\Lambda} = 0.6879$, $\Omega_{\mathrm{b}} = 0.0491$, $h = 0.6751$, $\sigma_8 = 0.8150$, and $n_{\mathrm{s}} = 0.9653$. All calibrations of the halo mass relations are performed on S\textsc{hark}.

\subsection{SAGE}
\label{subsec:SAGE}

The \textsc{Semi-Analytic Galaxy Evolution} (SAGE) model \citep{Croton2016} is a flexible, publicly available semi-analytic framework, building upon the Munich model lineage \citep{Croton2006}. SAGE incorporates detailed prescriptions for radiative cooling, star formation, stellar and AGN feedback, black hole growth, and environmental processes. The model parameters are tuned to match the observed stellar mass function and galaxy colour distributions at $z = 0$.

For this study, SAGE is applied to the BOLSHOI $N$-body simulation \citep{Klypin2011}, which covers a volume of $(250~h^{-1}~\mathrm{Mpc})^3$ with $2048^3$ particles, corresponding to a mass resolution of $1.35 \times 10^8~h^{-1}~M_{\odot}$. The SAGE simulated light cones used in this work consisted of 10 lightcones, each with an area of $\sim 1,960 \, \mathrm{deg}^2$. Haloes are identified using the \textsc{ROCKSTAR} phase-space halo finder \citep{Behroozi2013a}, and merger trees are constructed with the \textsc{Consistent Trees} algorithm \citep{Behroozi2013b}. A lower stellar mass limit of $\log M_{\star} > 7.5~[M_{\odot}]$ was applied to the mock light cones to ensure completeness and reliability in the resulting galaxy sample. The adopted cosmology is based on WMAP7 \citep{Komatsu2011}, with $\Omega_{\mathrm{m}} = 0.270$, $\Omega_{\Lambda} = 0.730$, $\Omega_{\mathrm{b}} = 0.0469$, $h = 0.70$, $\sigma_8 = 0.82$, and $n_{\mathrm{s}} = 0.95$. SAGE is used solely to validate the halo mass estimators without any re-tuning, thereby probing sensitivity to differing galaxy-formation prescriptions.

\subsection{GAEA}
\label{subsec:GAEA}

The \textsc{Galaxy Evolution and Assembly} (GAEA) semi-analytic model \citep{DeLucia2014, Hirschmann2016, DeLucia2024} provides an independent theoretical benchmark in this work. The latest version includes updated prescriptions for AGN feedback, environmental processes affecting satellites, black hole accretion, disk instabilities, and starburst activity \citep{Fontanot2020, DeLucia2011}.

GAEA parameters are calibrated to match the stellar mass function over $0 < z < 4$, local atomic and molecular hydrogen mass functions, and AGN bolometric luminosity function evolution to $z \sim 4$. The model is run on merger trees from the Millennium Simulation \citep{Springel2005}, which adopts a $\Lambda$CDM cosmology with $\Omega_{\mathrm{m}} = 0.25$, $\Omega_{\mathrm{b}} = 0.045$, $\Omega_{\Lambda} = 0.75$, $h = 0.73$, $n_{\mathrm{s}} = 1$, and $\sigma_8 = 0.8$. The simulation volume is $(500~h^{-1}~\mathrm{Mpc})^3$, with a particle mass of $8.625 \times 10^8~h^{-1}M_{\odot}$. The constructed observational cone from this simulation, was large enough to produce a full celestial sphere ($\sim 41,253 \, \mathrm{deg}^2$). A lower stellar mass limit of $\log M_{\star} > 8~[M_{\odot}]$ was applied to the mock light cones to ensure completeness and reliability in the resulting galaxy sample. Haloes and merger trees are constructed using \text{S}\textsc{ubfind} and \text{S}\textsc{ub}-LINK \citep{Springel2001}. GAEA is used solely to validate our halo mass estimates, without any re-tuning, thereby probing the estimates sensitivity to differing galaxy-formation prescriptions.

\section{Halo Mass Relations}
\label{sec:Methods}

In the group and cluster regime, halo mass estimates are traditionally derived from dynamical tracers, such as the velocity dispersion of member galaxies, or from abundance matching techniques that link observed galaxy properties to theoretical halo mass functions \citep{Yang2007, Viola2015, Lim2021}. However, these methods are subject to significant systematic uncertainties, particularly at low halo masses and for groups with low multiplicity, where the reliability of dynamical indicators is compromised by small number statistics and projection effects \citep{Old2015, Robotham2011, Muldrew2012}.

\begin{figure}[!hbt]
\centering
\includegraphics[width=\linewidth]{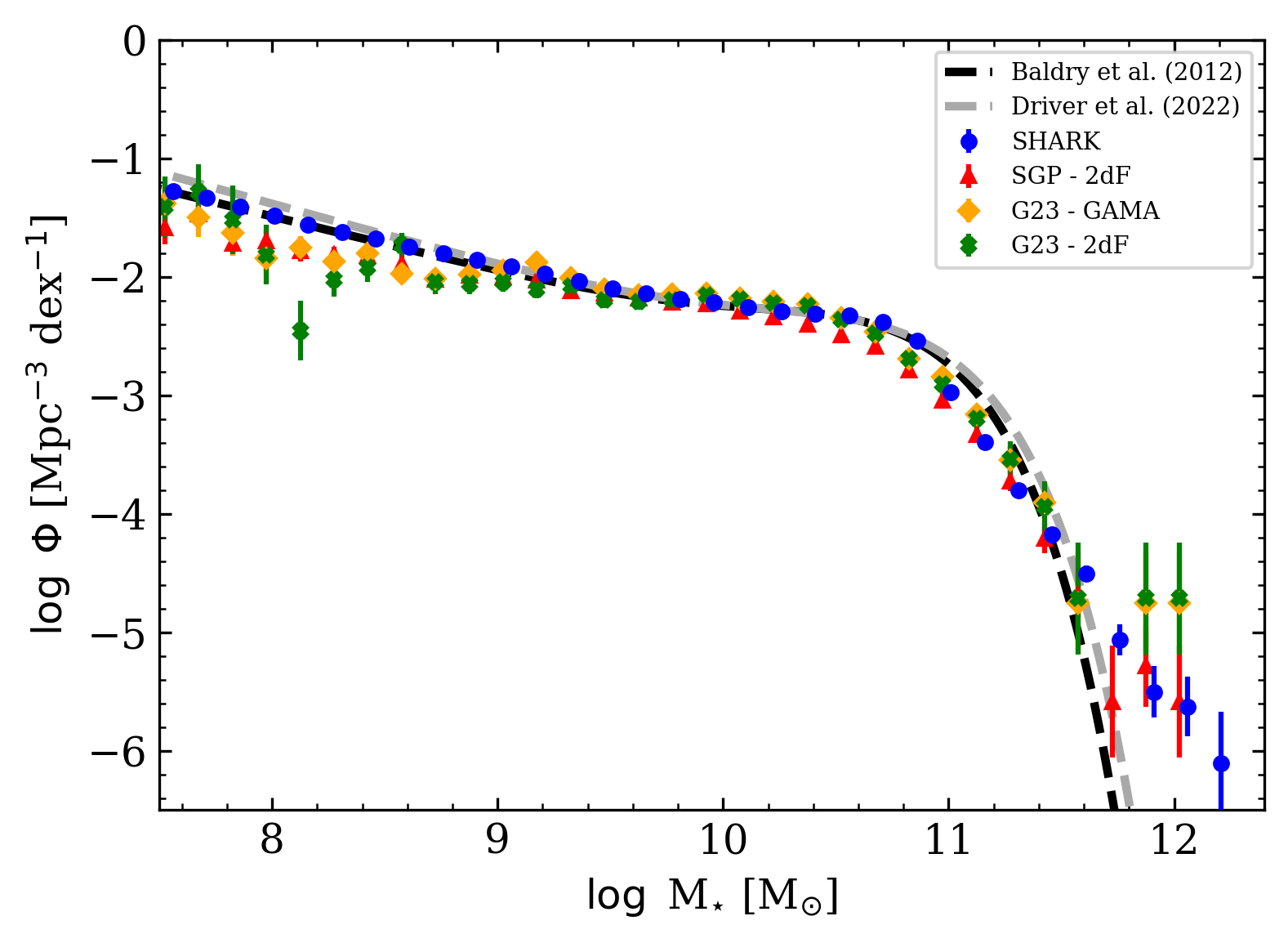}
\caption{Comparison of the SMF from S\textsc{hark} to well-established observational SMFs and SMFs produced by observational data from \cite{VanKempen2024}. The figure shows the SMF ($\log \Phi$) versus stellar mass ($\log M_{\star}$) for S\textsc{hark} v2.0 (blue circles) alongside canonical constraints from \citet{Baldry2012} (black dashed line) and \citet{Driver2022} (grey dashed line). Red triangles, orange diamonds, and green crosses represent our observational data from \citet{VanKempen2024} (SGP-2dF, G23-GAMA, and G23-2dF datasets, respectively). Error bars indicate Poisson uncertainties. S\textsc{hark} demonstrates excellent agreement with the established SMFs and across our observational datasets, with only minor deviations.}
\label{fig:SMF}
\end{figure}

\begin{figure}[!hbt]
\centering
\includegraphics[width=\linewidth]{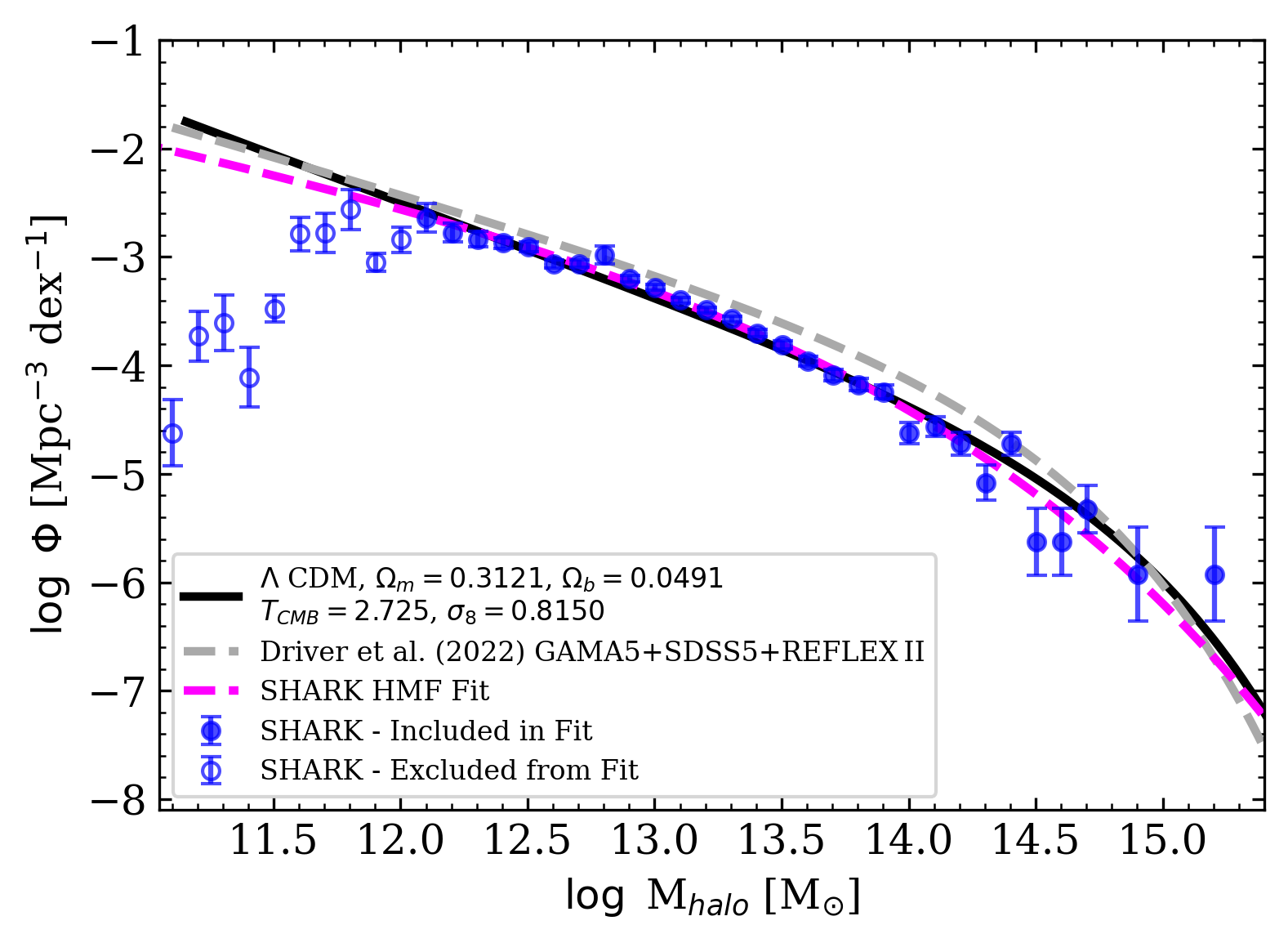}
\caption{Comparison of the HMF derived from S\textsc{hark} with analytic and observational benchmarks. The blue data points represent the S\textsc{hark} HMF, with error bars indicating Poisson uncertainties in each mass bin. The solid black curve denotes the analytic HMF prediction for the input cosmology of S\textsc{hark}, computed using the \texttt{hmf} Python package \citep{Murray2013}. The grey dashed line shows the empirical fit from \cite{Driver2022}, based on GAMA5, SDSS5, and REFLEX II data. The magenta dashed curve corresponds to a Schechter function fit to the S\textsc{hark} HMF.}
\label{fig:SHARK_HMF}
\end{figure}

To provide a physically motivated and observationally calibrated framework for halo mass estimation, we chose to calibrate our halo mass estimators to S\textsc{hark}. This decision is motivated by S\textsc{hark}'s demonstrated ability to reproduce key observables, most notably the observed stellar mass function (SMF; Figure~\ref{fig:SMF}) and halo mass function (HMF; Figure~\ref{fig:SHARK_HMF}), when measured using observational techniques. Furthermore, S\textsc{hark} is specifically tailored for application to forthcoming large-scale surveys, most notably WAVES, but also 4HS, and will serve as the primary simulation framework for training and assessing future group-finding algorithms to be used in these surveys \cite{Lagos2019}. In addition, recent advancements in S\textsc{hark}, such as the implementation of the HBT+HERONS merger tree algorithm \citep{ChandroGomez2025}, have significantly reduced numerical artefacts, including: mass swapping, massive transients, and orphan galaxies—that can arise in dark matter merger trees. These improvements yield a more stable and physically consistent halo population, which is essential for the development of robust and reliable halo mass estimators.

As shown in Figure~\ref{fig:SMF}, S\textsc{hark} provides an excellent match to well-established observational SMFs \citep{Baldry2012, Driver2022} across the full stellar mass range. The observational datasets from \citet{VanKempen2024} (SGP-2dF, G23-GAMA, and G23-2dF) show good agreement with the \cite{Baldry2012} and \cite{Driver2022} SMFs across the intermediate to high stellar mass range, which demonstrates the reliability of the sample in this regime. At lower stellar masses, however, the observed SMFs exhibit a deficit relative to these fits. This offset is attributable to the limitations of WISE photometry in detecting low-surface brightness and low stellar mass galaxies, resulting in systematic incompleteness at the faint end. Such incompleteness is a well-known limitation of near-infrared selected samples and must be considered when interpreting the low-mass behaviour of the observed SMF.

The SMF and HMF are constructed using a standard $1/V_{\mathrm{max}}$ approach to correct for survey incompleteness and selection effects. For a given mass bin ($\Delta \log M$; stellar mass for the SMF and halo mass for the HMF), the number density is computed as:
\begin{equation}
    \phi(\log M) = \sum_{i=0}^{N} \frac{1}{V_{\mathrm{max}}^i \, \Delta \log M},
\end{equation}
where $V_{\mathrm{max}}^i$ is the maximum comoving volume within which the $i$th galaxy (for the SMF) or group (for the HMF) could be observed, given the survey magnitude limits and selection criteria. For the SMF, $V_{\mathrm{max}}^i$ is determined by the redshift range over which each galaxy remains above the survey flux limit. For the HMF, following the methodology of \citet{Driver2022}, $V_{\mathrm{max}}^i$ is estimated based on the $n$th brightest galaxy in each group (here, $n=3$), such that the group's $V_{\mathrm{max}}$ is calculated based on the maximum survey volume that the $n$th brightest member remains above the survey flux limit. This approach ensures that the group sample is volume-limited with respect to its membership.

The halo mass function (HMF) is fit as a single Schechter function of the form:
\begin{equation}
\begin{array}{@{}l@{}}
\phi(\log M_{halo}) = \\ 
\hspace{3.5em}
    \ln(10) \, \phi_* \, \beta \left( \frac{M_{halo}}{M^*} \right)^{\alpha + 1}
    \cdot \exp \left[ - \left( \frac{M_{halo}}{M^*} \right)^{\beta} \right],
\end{array}
\end{equation}
where $M^*$ is the characteristic halo mass, $\phi_*$ is the normalisation, $\alpha$ is the low-mass slope, and $\beta$ is the exponential cutoff parameter. 

Figure~\ref{fig:SHARK_HMF} demonstrates the close agreement between the S\textsc{hark} HMF and analytic predictions, with only minor deviations at the lowest halo masses. The robust match between S\textsc{hark} and analytic expectations ensures that systematic biases in the halo mass distribution are minimised, providing a reliable foundation for calibrating observational mass proxies.

In the subsequent subsections, the formulation, calibration, accuracy, uncertainty and use cases of each halo mass estimator are described in detail. We employ SAGE and GAEA as independent tests to validate our S\textsc{hark}-calibrated relations. These models were run on different N-body simulations and contain different physical prescriptions, allowing us to assess whether our derived scaling relations remain robust across varied galaxy formation prescriptions and are suitable for general observational applications.

\subsection{Modifying The Virial Theorem}
\label{subsec:Meth1}

The evolution of dark matter haloes is governed solely by gravitational forces, a regime that is well explored through detailed N-body simulations and SAMs, whereas galaxies are complex systems regulated by a multitude of baryonic processes, including gas cooling, star formation, and feedback \citep[see review of ][]{Somerville2015b}. This fundamental distinction underpins the rationale for employing dynamical, gravity-based estimators for halo mass, which are expected to exhibit minimal dependence on the details of baryonic physics and thus provide a robust, model-independent approach to halo mass estimation.

%%%%%%%%%%%%%%%%%%%%%%%%%%%%%%%%%%%%%%%%%%%%%%%%%
\subsubsection{The Virial Theorem}
\label{subsubsec:VT_VT}

\begin{figure*}[!hbt]
\centering
\includegraphics[width=\textwidth]
{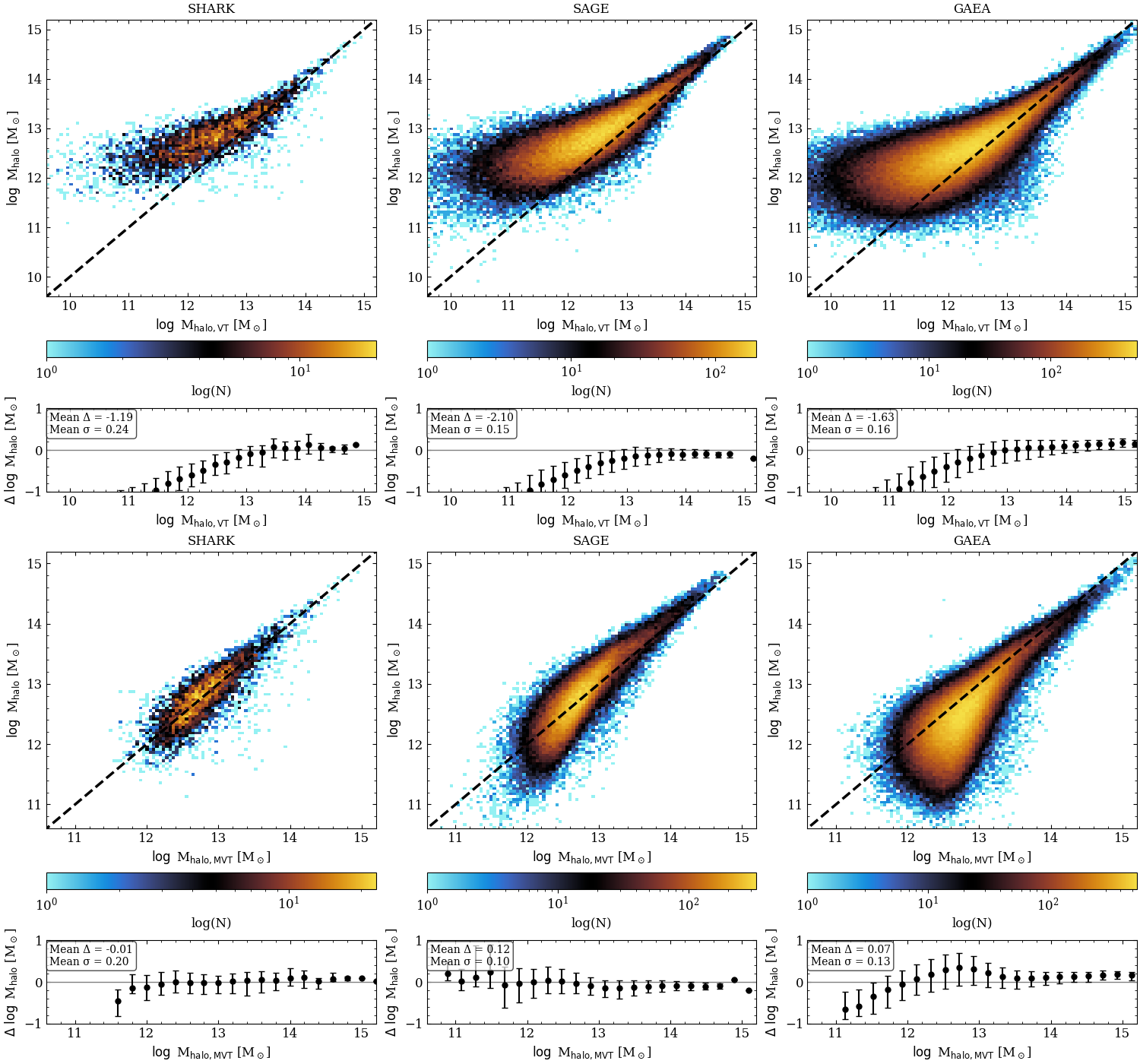}
 \caption{ Comparison of halo mass estimates derived from the velocity dispersion relation before and after calibration, across three semi-analytic models: S\textsc{hark}, S\textsc{AGE}, and G\textsc{AEA}. All samples have a manitude limit of $i<19.2$ and a redshift limit of $z<0.1$. \textbf{Top Row:} The virial theorem halo mass ($\log M_{\mathrm{halo, \, VT}}$) versus the true halo mass ($\log M_{\mathrm{halo}}$) for each model, with colour indicating the logarithmic number density of halos in each bin. The dashed line denotes the one-to-one relation. \textbf{Second Row:} Corresponding residuals ($\Delta \log M_{\mathrm{halo}} = \log M_{\mathrm{halo, \, VT}} - \log M_{\mathrm{halo}}$) and error bars indicate $16^{th}$ and $84^{th}$ percentiles in each bin. The mean offset and scatter indicated in the legend. \textbf{Third Row:} The MVT halo mass estimates ($\log$ M$_{\mathrm{halo, \, MVT}}$), compared to the true halo mass ($\log M_{\mathrm{halo}}$). \textbf{Bottom Row:} Corresponding median of residuals ($\Delta \log M_{\mathrm{halo}} = \log M_{\mathrm{halo, \, MVT}} - \log M_{\mathrm{halo}}$) and error bars indicate $16^{th}$ and $84^{th}$ percentiles in each bin. The mean offset and scatter indicated in the legend.
 }
    \label{fig:Disp_HM}
\end{figure*}

In observational studies of galaxy groups and clusters, the velocity dispersion is measured along the line of sight, denoted as $\sigma_{\mathrm{los}}$ (in units of km s$^{-1}$). For a self-gravitating, isotropic, and uniform sphere of mass $M$ and radius $R$, the virial theorem relates the total kinetic energy $T$ and potential energy $U$ as $2T + U = 0$ in equilibrium. The total kinetic energy can be expressed in terms of $\sigma_{\mathrm{los}}$ as $T = \frac{1}{2} M \sigma_{\mathrm{los}}^2$, while the gravitational potential energy remains $U = -\frac{3}{5} \frac{GM^2}{R}$. This yields the standard virial mass estimator:
\begin{equation}
    M_{\mathrm{halo,~VT}}(h^{-1}M_{\odot}) = \frac{5}{3} \frac{\sigma_{\mathrm{los}}^2 R}{G},
\label{Eq:OG_VT}
\end{equation}
where $R$ is a radius (e.g., the maximum projected separation among group members from the median centre of the group) in units of $\mathrm{Mpc}$, and $G$ is the gravitational constant in units of $M_{\odot}^{-1}\,\mathrm{km}^2\,\mathrm{s}^{-2}\,\mathrm{Mpc}$, yielding a mass in $h^{-1}M_{\odot}$. This estimator is widely used in both observational and theoretical studies of galaxy groups and clusters \citep[e.g.,][]{Carlberg1996, Eke2004, Robotham2011}. While the $\frac{5}{3}$ coefficient is sometimes omitted in the literature \citep[e.g.][]{Finn05, Evrard2008, Lau2010, Poggianti2010, Robotham2011, Alpaslan2012}, it is physically motivated and essential for accurate mass estimates, particularly for massive halos.

The line-of-sight velocity dispersion, $\sigma_{\mathrm{los}}$, is calculated using the Gapper (gap) method \citep{Beers1990}, which is particularly robust for small group sizes and is less sensitive to outliers than standard deviation-based estimators. For a group of $N$ galaxies with ordered velocities $v_1 < v_2 < \ldots < v_N$, the Gapper estimator is defined as:
\begin{equation}
\sigma_{\mathrm{gap}} = \frac{\sqrt{\pi}}{N(N-1)} \sum_{i=1}^{N-1} w_i g_i,
\end{equation}
where $g_i = v_{i+1} - v_i$ is the velocity gap between adjacent ordered velocities, and $w_i = i(N-i)$ is a weighting factor. Through the use of ordered velocity gaps, the gap method provides lower sampling variance than standard-deviation or bi-weight scale estimators, and is less sensitive to interlopers and has thus become a standard approach in modern group catalogues \citep[e.g.,][]{Eke2004, Robotham2011, Old2015}.

In most mock haloes, the brightest galaxy is assumed to be at rest with respect to the halo centre of mass. To account for this, the velocity dispersion is further corrected by a factor of $\sqrt{N/(N-1)}$ \citep{Eke2004, Robotham2011}. The final velocity dispersion is then given by:
\begin{equation}
\sigma = \sqrt{\frac{N}{N-1} \sigma_{\mathrm{gap}}^2 } .
\end{equation}
This approach ensures that the velocity dispersion estimates are unbiased and robust, even for low-multiplicity systems, and that the dominant sources of observational uncertainty are properly propagated \citep{Robotham2011}.

The top panels of Figure~\ref{fig:Disp_HM} compare the true halo masses from the simulations to those measured using the virial theorem via Equation~\ref{Eq:OG_VT} on the simulated light-cones with a $z<0.1$ limit. These results demonstrate that the traditional virial theorem mass estimator is subject to substantial scatter and systematic offsets, particularly at low halo masses and for systems with low multiplicity. As shown in the lower panels, the mean offset ($\Delta$) between the estimated and true halo masses is significant across all models, reaching values as large as $-2.10$ dex (SAGE), $-1.63$ dex (GAEA), and $-1.19$ dex (SHARK). These biases indicate that, without modification, the virial theorem systematically underestimates halo masses, especially in the low-mass regime, highlighting the need for improved halo mass estimates in group environments. Given these limitations, it is essential to develop improved approaches for halo mass estimation in group environments. In the following section, we present a calibrated virial theorem estimator designed to address these shortcomings and provide more reliable halo mass measurements.

%%%%%%%%%%%%%%%%%%%%%%%%%%%%%%%%%%%%%%%%%%%%%%%%%%%%
\subsubsection{Calibration of the Modified Virial Theorem}
\label{subsubsec:VT_MCMC}

While the virial theorem provides a physically motivated starting point, its direct application to observed galaxy groups is complicated by departures from equilibrium, projection effects, and uncertainties in group membership, especially for low-multiplicity systems \citep{Robotham2011, Muldrew2012, Old2015}. To account for these effects, we introduce a modified virial theorem (MVT), which includes a scale factor, $A$, to Equation~\ref{Eq:OG_VT} and corrects the mass estimate based on the measured velocity dispersion and projected radius of the group:
\begin{equation}
    M_{\mathrm{halo,~MVT}} = A \frac{\sigma^2 R}{G},
\label{Eq:Cor_VT}
\end{equation}
where $A$ is a function of both velocity dispersion and group radius. The calibration of $A$ is essential to ensure unbiased halo mass estimates across the full range of group halo masses.

The calibration of the MVT was performed using a Bayesian framework, employing Markov Chain Monte Carlo (MCMC) sampling to explore the posterior probability distribution of the model parameters. The likelihood function was constructed to minimise the scatter in $\log M_{\mathrm{halo}}~[M_{\odot}]$ at fixed predicted halo mass, effectively minimising $\chi^2$ and maximising the log-likelihood. Convergence of the MCMC chains was assessed via autocorrelation analysis. Further details of the likelihood function and the Bayesian inference methodology are provided in \ref{sec:VT Fitting}.

A suite of functional forms for the correction terms $A_{\sigma}$ and $A_{R}$ were tested, including exponential, inverse square, linear, logistic, and sigmoid decay models. The power-law form was found to provide the best fit to the simulation data, as quantified by the Akaike Information Criterion (AIC) and Bayesian Information Criterion (BIC), outperforming alternative models by a substantial margin. The final adopted form for the $A$ coefficient is:
\begin{equation}
    A= \frac{5}{3} + A_{\sigma} + A_{R},
\end{equation}
with the power-law decay models for $A_{\sigma}$ and $A_{R}$ given by:
\begin{equation}
    A_{\sigma} = 
    \begin{cases}
    \alpha \left( \left( \frac{ \sigma }{ \sigma_{lim} } \right)^{n_1} - 1 \right), & \text{if } \sigma \leq \sigma_{lim}\\
    0, & \text{otherwise}
    \end{cases},
\label{Eq:Sig_lim}
\end{equation}
\begin{equation}
    A_{R} = 
    \begin{cases}
    \beta \left( \left( \frac{ R }{ R_{lim} } \right)^{n_2} - 1 \right), & \text{if } R \leq R_{lim}\\
    0, & \text{otherwise}
    \end{cases},
\label{Eq:R_lim}
\end{equation}
where $\alpha$, $\sigma_{lim}$, $n_1$, $\beta$, $R_{lim}$, and $n_2$ are free parameters determined by the MCMC sampling. The resulting posterior distributions for all model parameters are presented in Figure~\ref{fig:MCMC_Disp} in \ref{sec:VT Fitting}, demonstrating that the parameters are well-constrained. The best-fitting parameters for the MVT are given in Table \ref{Tab:Disp_Coeff} for convenience.

\begin{table}[!htb]
\caption{Best-fitting parameters for the calibrated virial theorem relation.}
\centering
\begin{tabularx}{\columnwidth}{ CCCCCC }
\hline \hline 
$\alpha$  &  $\sigma_{lim}[\mathrm{km/s}]$ &  $n_{1}$  &   $\beta$   &  $R_{lim}[\mathrm{Mpc}]$   &  $n_{2}$\\ 
\hline
1.030     &  244.634                   &   -1.989  &   0.213     &    0.369          &  -1.591\\
\hline \hline
\end{tabularx}
\label{Tab:Disp_Coeff}
\end{table}

Notably, the calibrated $A_{\sigma}$ and $A_{R}$ coefficients decay to zero at the fitted velocity dispersion and group radius limits, reflecting that for larger multiplicity groups, the velocity dispersion and radius are well-defined and no further modification is required. Above these limits, the MVT (Equation~\ref{Eq:Cor_VT}) reduces to the virial theorem (Equation~\ref{Eq:OG_VT}). This procedure ensures that the MVT estimator is unbiased and robust across the full range of group properties, with well-quantified uncertainties and minimal model dependence.

To convert halo masses between different values of the Hubble parameter $H_0$ (or Hubble constant $h$), the following scaling should be applied:
\begin{equation}
    R_{lim}^{'} = R_{lim} \cdot \frac{h}{h^{'}},
\end{equation}
where $R_{lim}^{'}$ is the group radius limit for a newly chosen $h'$, and $h$ is the value used in this work ($h=0.7$). For example, converting to $h'=0.67$ yields $R_{lim}^{'} =  0.369\cdot \frac{0.7}{0.67}=0.386$ Mpc. Alternatively, a factor of $\log_{10} (h'/h)$ can be added to the calculated $M_{\mathrm{halo}}$ from Equation~\ref{Eq:Cor_VT} when using the $R_{lim}$ in Table~\ref{Tab:Disp_Coeff}. The calculated projected group radius for a given group should also be recalculated for the chosen $h$ value.

\subsubsection{MVT - Accuracy, Uncertainty and Use Cases}
\label{subsubsec:VT_ACC_Unc}

The MVT is calibrated to S\textsc{hark}, we then quantify the calibration accuracy and uncertainty on S\textsc{hark} and cross-validate the calibrated estimator to both SAGE and GAEA. The bottom panels of Figure~\ref{fig:Disp_HM} present a quantitative comparison between the predicted halo mass from the MVT (Equation \ref{Eq:Cor_VT}) and true halo masses for each model (calibrated to $z < 0.1$ and $i<19.2$ mag). The accuracy of the MVT is characterised by the mean offset (mean $\Delta$) between the predicted and true halo masses, while the uncertainty is quantified by the mean of the $16^{\mathrm{th}}$ and $84^{\mathrm{th}}$ percentiles (mean $\sigma$) of the residuals. These metrics are shown in the bottom panels of Figure~\ref{fig:Disp_HM} as a function of the predicted halo mass for each simulation. On the calibration model (S\textsc{hark}), the mean $\Delta$ is $-0.01$ dex, with a mean $\sigma$ of $0.20$ dex, indicating negligible systematic bias and moderate scatter. On the validation models; SAGE yields a mean $\Delta$ of $0.12$ dex and a mean $\sigma$ of $0.10$ dex, while the GAEA model exhibits a mean $\Delta$ of $0.07$ dex and a mean $\sigma$ of $0.13$ dex. These results demonstrate that the calibration procedure effectively removes systematic biases and reduces scatter across the full mass range and for all group multiplicities. The calibrated relation yields a high degree of consistency in both normalisation and slope across the models compared to the traditional virial theorem estimator (top panels), with only minor differences from the true halo mass, particularly at the low-mass end in the GAEA model. The small variance in the calibrated relation between the different models reflects the minimal dependence of the virial theorem on the details of baryonic physics, as it is fundamentally anchored in gravitational dynamics. However, some residual differences are observed, particularly in the low-mass regime for the GAEA simulation.

The robustness of the calibrated relation was further tested by varying the magnitude and redshift limits. When imposing a fainter WAVES-wide magnitude limit ($Z < 21.2$), the free parameters changed by 1–20\%, with the correction factor decreasing due to the deeper limiting magnitude. Extending the redshift range to $z < 0.3$ resulted in larger changes of 5–40\% in the free parameters, as stronger corrections were required as a function of redshift. These tests demonstrate that the calibrated relation presented here is specifically tailored to GAMA-like selection criteria (i.e., $i < 19.2$) and to $z < 0.1$. Users applying this relation to datasets with different selection functions or redshift limits should be aware of the potential systematic biases that may arise when doing so and a recalibration of the relation for their specific survey parameters is suggested. Nevertheless, despite these limitations, the application of the calibrated relation to non-ideal datasets remains preferable to relying solely on the traditional virial theorem, which does not account for selection effects or redshift-dependent biases.

In summary, the MVT is the least model-dependent method considered in this work, as it does not require knowledge of galaxy properties or detailed prescriptions for baryonic processes amongst the models. This makes it ideally suited for applications where unbiased halo mass estimates are required, such as the construction of the halo mass function (HMF) in spectroscopic surveys or the derivation of cosmological parameters from group catalogues, provided that the relation is applied to the appropriate dataset.

\subsection{The summed Stellar-to-Halo Mass Relation}
\label{subsec:Meth2}

\begin{figure*}[!hbt]
\centering
\includegraphics[width=\linewidth]{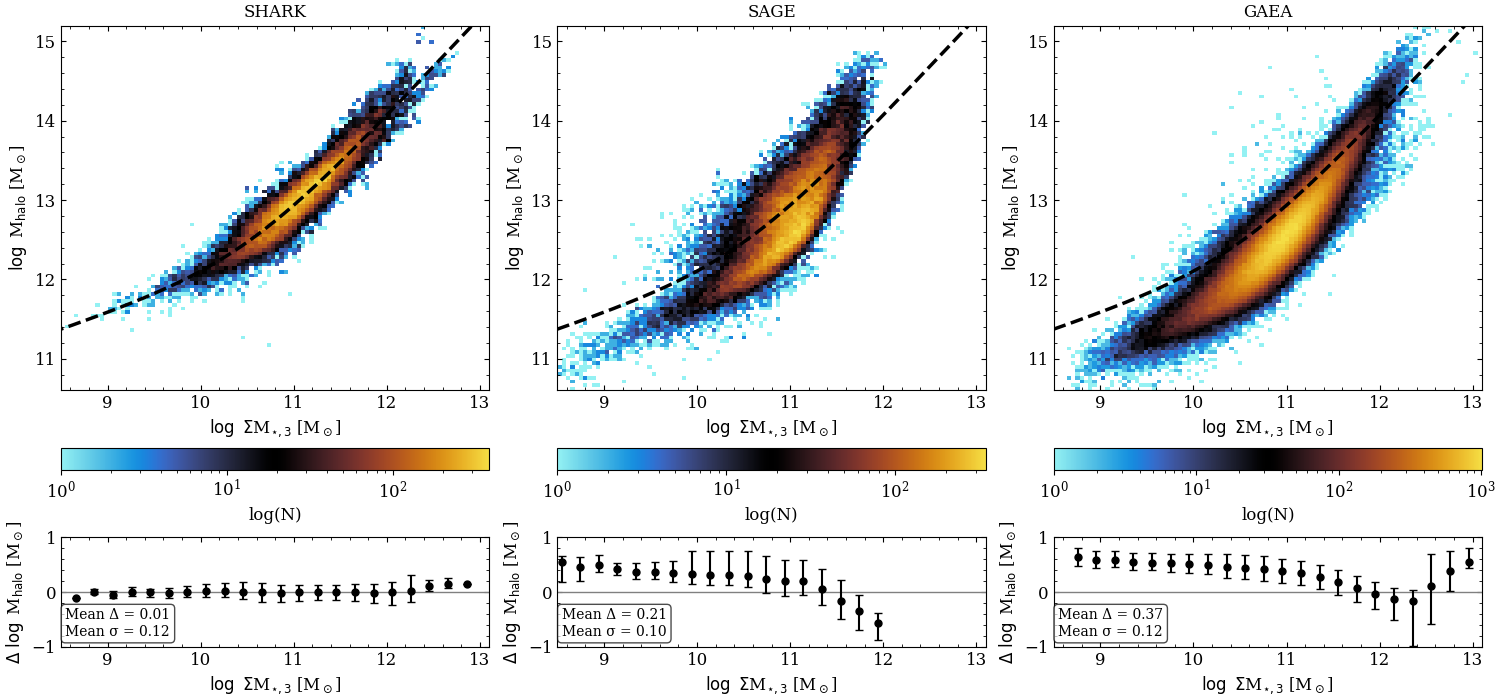}
\includegraphics[width=\linewidth]{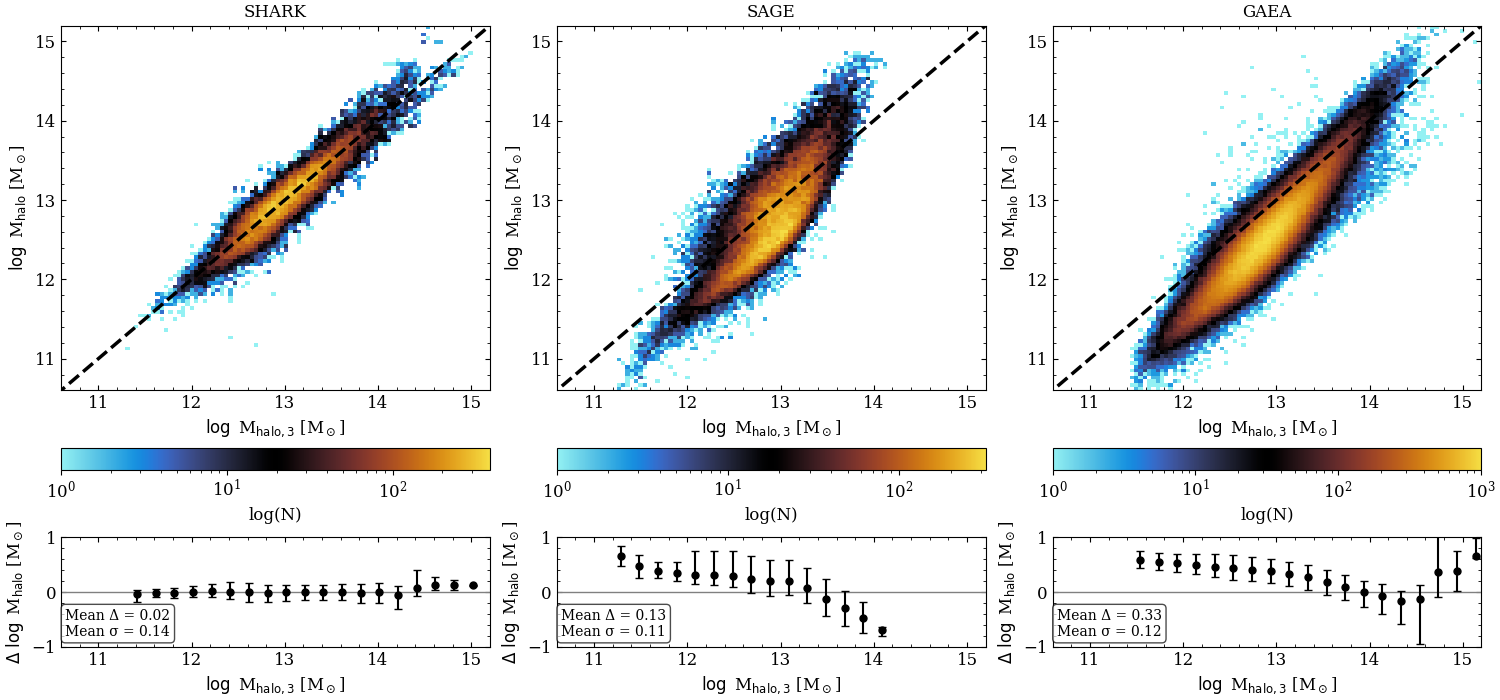}
\caption{
Comparison of baryonic and halo mass relations across three SAMs: S\textsc{hark}, SAGE, and GAEA. The S\textsc{hark} sample has a fainter magnitude limit of $Z<21.2$ and a deeper redshift limit of $z<0.3$ compared to SAGE and GAEA ($i<19.2$ and $z<0.1$) \textbf{Top Row:} The sSHMR using the summed stellar mass of the three most massive group galaxies (\(\log \Sigma M_{*, \mathrm{3}}\)) and the true halo mass (\(\log M_{\mathrm{halo}}\)) is shown for each SAM, with the colour scale indicating the logarithm of the number of groups per bin. The black dashed line in each panel represents the MCMC-derived fit to the S\textsc{hark} data, optimised to minimise scatter in halo mass. This same fit is overlaid on the SAGE and GAEA panels to illustrate model dependence. \textbf{Second Row:} The residuals in halo mass (\(\Delta \log M_{halo} = \log \Sigma M_{*, \, \mathrm{3}} - \log M_{\mathrm{halo}}\)) is shown as a function of the summed stellar mass. \textbf{Third Row:} Estimated halo masses, derived by applying the S\textsc{hark}-calibrated sSHMR, are plotted against the true halo masses from each simulation. The black dashed line denotes the one-to-one relation. The close alignment of the points along this line in all models demonstrates that the S\textsc{hark}-based calibration provides robust halo mass estimates, with low bias and scatter, even when applied to independent SAMs. \textbf{Bottom Row:} The residuals in halo mass are shown as a function of the estimated halo mass. In both residual panels, the colour scale again indicates the logarithm of the number of groups per bin.
}
\label{fig:SHMR}
\end{figure*}

The connection between the stellar content of galaxies and their host dark matter haloes is a cornerstone of galaxy formation theory, providing a critical link between observable baryonic properties and the underlying dark matter distribution \citep{Behroozi2010, Moster2013, Wechsler2018}. While abundance matching and halo occupation models have traditionally been used to infer this relationship on a statistical basis \citep[e.g.,][]{Yang2003, Vale2004, Behroozi2010, Moster2013, Kravtsov2018}, direct group-based approaches offer a powerful means to calibrate halo mass estimators for individual systems \citep[e.g.,][]{Viola2015, Lim2021}, particularly in the group regime where dynamical methods become less reliable at low multiplicity \citep{Old2015, Muldrew2012}.

%%%%%%%%%%%%%%%%%%%%%%%%%%%%%%%%%%%%%%%%%%%%%%%%%%%%%%%%%%%%%%%%%%%%%%%%%%%%
\subsubsection{Summed Stellar Mass Proxy}
\label{subsubsec:SHMR_Proxy}

In this section, we establish an empirical sSHMR between the summed stellar masses of the three most massive galaxies in a group ($\Sigma M_{*,\mathrm{3}}$) and the group halo mass ($M_{\mathrm{halo}}$). This approach is motivated by the expectation that the most massive group members should provide the most reliable baryonic tracers of the underlying halo mass. By summing the stellar mass of the three most massive galaxies, this method should mitigate stochasticity compared to a typical single galaxy tracer via a stellar-to-halo mass relation (SHMR). The choice of three galaxies represents a balance between competing factors, using only the most massive galaxy may introduce excessive scatter and result in a steeper high-mass slope in the SHMR, where small changes in stellar mass would produce disproportionately large changes in the estimated halo mass. Conversely, extending the sum to more galaxies may yield diminishing returns while adding noise from less reliable tracers. Summing over the three most massive galaxies should flatten the relation at the high-mass end and provide a more stable and physically motivated estimator. Additionally, we assess the robustness of this relation by varying the selection criteria from the GAMA-like limit ($i < 19.2$) to a fainter WAVES-wide limit ($Z < 21.2$) and by extending the redshift range from $z < 0.1$ to $z < 0.3$, examining how the relation responds to different survey parameters.

%%%%%%%%%%%%%%%%%%%%%%%%%%%%%%%%%%%%%%%%%%%%%%%%%%%%%%%%%%%%%%%%%%%%%%%%%%%%
\subsubsection{sSHMR Calibration and Functional Form}
\label{subsubsec:SHMR_MCMC}

We use the same MCMC methodology as in Section~\ref{subsubsec:VT_MCMC}. The likelihood function was constructed to minimise the scatter in $\log$ $M_{\mathrm{halo}}$ at fixed $\log \Sigma M_{\star, \, 3}$; convergence was assessed via autocorrelation analysis. The resulting posterior distributions for the model parameters are shown in Figure~\ref{fig:MCMC_P3M}, and further details of the likelihood function, the fitting methodology and the uncertainty of the fitted parameters are provided in \ref{sec:SHMR Fitting}.

The adopted functional form for the sSHMR is a double power-law in the form:
\begin{equation}
\begin{array}{@{}l@{}}
M_{\mathrm{halo,\, 3}} \,(M_{\odot}) = \\
\hspace{3.5em}
A \cdot \Sigma M_{*,\, \mathrm{3}} \cdot  
\left( \left( \frac{\Sigma M_{*,\, \mathrm{3}}}{M_A} \right)^{\beta} 
+ \left( \frac{\Sigma M_{*,\, \mathrm{3}}}{M_A} \right)^{\gamma} \right),
\end{array}
\label{Eq:SHMR}
\end{equation}
where $\Sigma M_{*,\mathrm{3}}$ is the sum of the stellar masses of the three most massive galaxies in the group in units of $M_{\odot}$, $A$ is the normalisation, $M_A$ is the characteristic stellar mass in units of $M_{\odot}$, and $\beta$ and $\gamma$ are the low- and high-mass slopes, respectively. The best-fitting parameters from the MCMC sampling are summarised in Table~\ref{Tab:SHMR_Coeff}.

\begin{table}[!htb]
\caption{Best-fitting parameters for the summed stellar mass–halo mass relation.}
\centering
\begin{tabularx}{\columnwidth}{ CCCC }
\hline \hline 
$A$  &  $\log M_A[M_{\odot}]$  &  $\beta$  &  $\gamma$ \\ 
\hline
46.944 & 10.483 & 0.249 & -0.601 \\
\hline \hline
\end{tabularx}
\label{Tab:SHMR_Coeff}
\end{table}

To convert the SHMR between different values of the Hubble parameter $H_0$ (or h), the following scaling should be applied:

\begin{equation}
    M_{A}^{'} = M_A \cdot \frac{h}{h^{'}},
\end{equation}
where $M_{A}^{'}$ is the characteristic stellar mass for a newly chosen $h^{'}$, and h is the value used in this work ($h=0.7$). For example, converting to $h^{'} = 0.67$ yields $\log M_{A}^{'} = \log (10^{10.483} \cdot \frac{0.7}{0.67}) = 10.502$. As the characteristic stellar mass is derived from simulations, there is only one factor of $h^{-1}$ rather than the typical $h^{-2}$ for observations. It is essential to ensure that both the stellar masses used in the sSHMR and the characteristic mass parameter $M_A$ are consistently defined with respect to the chosen value of $h$. Any conversion of the sSHMR to a different Hubble parameter must be accompanied by a corresponding adjustment of these quantities.

%%%%%%%%%%%%%%%%%%%%%%%%%%%%%%%%%%%%%%%%%%%%%%%%%%%%%%%%%%%%%%%%%%%%%%%%%%%%
\subsubsection{sSHMR - Accuracy, Uncertainty and Use Cases}
\label{subsubsec:SHMR_Acc_Unc}

Following the same methods as in Section~\ref{subsubsec:VT_ACC_Unc}, the sSHMR is calibrated to S\textsc{hark} where we test the accuracy and uncertainty of this calibration, then we-cross validate this with SAGE and GAEA as independent tests. The accuracy of the sSHMR calibration is characterised by the mean $\Delta$ between the predicted and true halo masses, while the uncertainty is quantified by the mean $\sigma$. Both are presented in the second and last rows of Figure~\ref{fig:SHMR} as a function of both the sum of the primary three stellar masses ($\Sigma M_{\star,\,3}$) and the predicted halo mass ($M_{\mathrm{halo,\,3}}$) for each model. On the calibration model (S\textsc{hark}), the mean $\Delta$ is $0.01$ dex as a function of the summed masses, with a mean $\sigma$ of $0.12$ dex; as a function of the predicted halo mass, the mean $\Delta$ is $0.02$ dex with a mean $\sigma$ of $0.14$ dex, indicating negligible systematic bias. Compared to the MVT in Section~\ref{subsubsec:VT_ACC_Unc}, the sSHMR achieves similar accuracy but with approximately half the uncertainty in the S\textsc{hark} model. 

Through cross-validation the SAGE and GAEA models yield mean $\Delta$ values of $0.21$ dex and $0.37$ dex, respectively, and mean $\sigma$ values of $0.10$ dex and $0.12$ dex, respectively, as a function of the summed stellar masses. As a function of the predicted halo mass, SAGE and GAEA yield mean $\Delta$ values of $0.13$ dex and $0.33$ dex, respectively, and mean $\sigma$ values of $0.11$ dex and $0.12$ dex, respectively. The differences in the sSHMR between models are more pronounced than those seen for the MVT. These discrepancies arise from the varying treatments of star formation efficiency, feedback, and merger-driven stellar mass growth in each SAM, as well as differences in the underlying dark matter merger trees and halo assembly histories. Such model dependencies are an inherent limitation of baryonic tracers, as the stellar mass content of halos is sensitive to the adopted physical prescriptions and calibration strategies.

As hypothesised, using a sSHMR proved to be the most effective approach rather than a typical SHMR. This method reduces the impact of stochasticity compared to using a single galaxy tracer, while also avoiding the diminishing returns observed when incorporating more than three members. Importantly, extending the relation to include more than three galaxies would, by definition, exclude groups with fewer than four members from being appropriately sampled. Such groups would require a separate relation, introducing additional complexity and reducing the universality of the method. Relying solely on the most massive galaxy leads to a steep slope at the high-mass end, where small increases in stellar mass correspond to disproportionately large increases in halo mass. In contrast, the use of $\Sigma M_{\star,\,3}$ flattens this relation, producing a more gradual and stable increase in halo mass with stellar mass. This improvement enhances its suitability for observational studies, as uncertainties in stellar mass estimates translate to significantly smaller errors in the inferred halo mass. The relation between $\Sigma M_{\star,\,3}$ and the true halo mass exhibits a high degree of consistency in overall form, but with notable differences in normalisation and slope between the models, as shown in Figure~\ref{fig:SHMR}. These differences reflect the distinct implementations of baryonic physics, feedback, and merger histories in each SAM, which influence both stellar mass assembly and the mapping between baryonic and dark matter components.

Testing the robustness of the relation across different selection criteria demonstrated that extending the redshift range to $z < 0.3$ produced no significant changes, with the fitted free parameters varying by less than 5\%. The relation shown in Figure~\ref{fig:SHMR} is calibrated on S\textsc{hark} using the fainter WAVES-wide limit ($Z < 21.2$) extended to $z < 0.3$. This is what is shown within the S\textsc{hark} panels of Figure~\ref{fig:SHMR}, whilst SAGE and GAEA both display a sample with ($i < 19.2$) and $z < 0.1$. Importantly, using the brighter GAMA-like limit ($i < 19.2$) fails to capture the power-law slope at the low-mass end, making it unsuitable for calibrating this relation. The fainter $Z < 21.2$ limit captures the low-mass regime, enabling a complete characterisation of the sSHMR. Crucially, while the relation itself does not depend on the magnitude limit, the ability to define this relation does. This means the calibrated relation can be applied to any selection function with a brighter limit than $Z < 21.2$, making it applicable to most existing group catalogues. Furthermore, since this method uses stellar mass rather than luminosity as the mass tracer, it can be applied to any group catalogue where reliable stellar mass measurements are available.

It is important to note that, while the sSHMR provides the most precise and lowest-scatter halo mass estimates, it is also the most model-dependent of the methods considered. The distribution of halo masses is tied to both the HMF and SMF of the input model—S\textsc{hark}. As a result, this halo mass estimator is not suitable for observationally constructing the HMF or for applications that aim to produce unbiased cosmological parameters from halo masses. For such applications, the MVT method is preferred, as its model dependence is negligible. Nevertheless, for studies focused on group-scale halo mass estimation, where the primary goal is to minimise residuals and uncertainties and to obtain a precise halo mass measurement, the sSHMR is the optimal choice, provided that reliable stellar mass measurements are available.

In summary, the calibrated sSHMR offers a powerful and practical tool for estimating group halo masses, delivering the highest precision among the methods tested. Its application should be restricted to contexts where model dependence is not a limiting factor. The method's versatility is demonstrated by its applicability to any group catalogue with stellar mass measurements and its robustness across different survey selection functions, provided they have magnitude limits brighter than $Z < 21.2$. Critically, users must ensure that their stellar mass measurements are analogous to ProSpect-derived stellar masses, as the S\textsc{hark} model used for this calibration is specifically tailored to ProSpect stellar masses. Failure to account for any systematic differences between stellar masses may introduce biases in the resulting halo mass estimates, particularly at the high-mass end.

\section{Observational Application of Halo Masses}
\label{sec:ObsApp}

The calibrated halo mass estimators developed in this work enable a range of new applications in the analysis of spectroscopic group catalogues. Here, we present two representative examples using the SGP group sample from \cite{VanKempen2024}: (i) the construction of an empirical HMF using the MVT, and (ii) the mapping of the quenched fraction of galaxies in the stellar mass–halo mass plane using the sSHMR.

%%%%%%%%%%%%%%%%%%%%%%%%%%%%%%%%%%%%%%%%%%%%%%%%%%%%%%%%%%%%%%%%%%%%%%%%%%%%%%%%%%%%%%%%%%%%

\subsection{Halo Mass Function from Observational Groups}
\label{subsec:Obs_HMF}

\begin{figure*}[t]
\centering

\includegraphics[width=\textwidth]{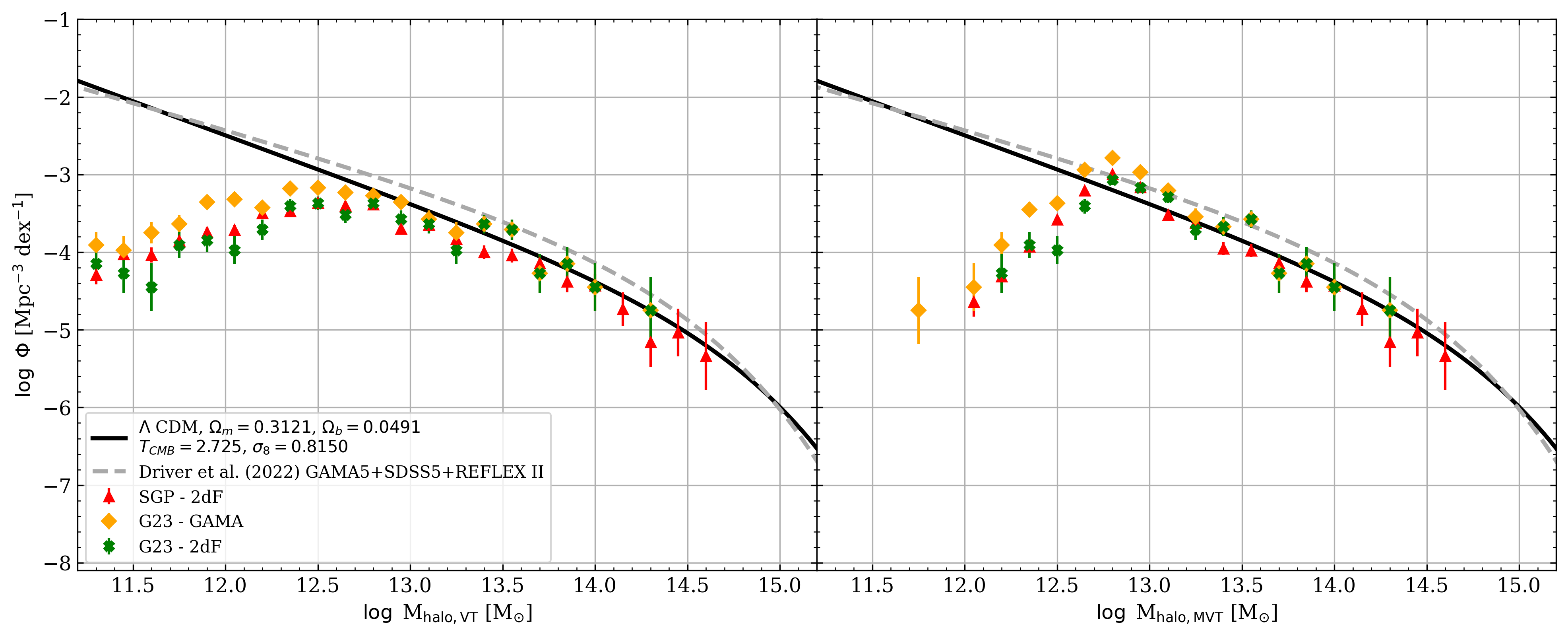}
\caption{Empirical HMF for galaxy groups with three or more members, constructed from the group catalogue of \citet{VanKempen2024}. The HMF is shown separately for groups in the SGP region dominated by 2dF coverage (red triangles), the G23 region with GAMA spectroscopy (orange diamonds), and the G23 region with 2dF spectroscopy (green crosses). Halo masses are estimated using the traditional and calibrated virial theorem method ($\log M_{\mathrm{halo,\,VT}}$ or $\log M_{\mathrm{halo,\,MVT}}$). Number densities are not corrected for survey volume or selection effects, due to the heterogeneous nature of the group sample and the challenges in defining a complete selection function. The solid black curve represents the analytic HMF prediction for the adopted cosmology of S\textsc{hark}, while the grey dashed line shows the empirical fit from \citet{Driver2022} based on GAMA5, SDSS5, and REFLEX II data. Error bars reflect Poisson uncertainties.}
\label{fig:Obs_HMF}
\end{figure*}

The HMF is a fundamental cosmological observable, that encodes information about the growth of structure and the underlying cosmological parameters \citep[e.g.,][]{Press1974, Jenkins2001, Murray2013}. In Figure~\ref{fig:Obs_HMF}, we present the empirical HMF for galaxy groups with $N_{\mathrm{m}} \geq 4$ members. The sample is divided into three subsamples: the SGP-2dF (red triangles) which uses 2dFGRS photometry across the entire sample, the G23-2dF region (green crosses) which uses 2dFGRS photometry in the G23 region, and the G23-GAMA (orange diamonds) which uses the GAMA photometry.

 Unlike the SMF, which is based on uniform measurements of individual galaxies, constructing the HMF requires consistent data for all members within each group and a well-defined selection function to enable $1/V_{\mathrm{max}}$ corrections. However, the observational group sample is heterogeneous, with the formation and detection of groups often drawing from multiple spectroscopic surveys that have varying completeness and magnitude limits \citep[see][]{VanKempen2024}. This diversity makes it challenging to accurately determine the completeness for each group, and thus we do not apply the $1/V_{\mathrm{max}}$ correction to the observational HMF. The absence of this correction means that the HMF is not fully corrected for incompleteness, particularly at the low-mass end. Low-mass groups are more likely to be missed because their member galaxies often fall below the survey’s limiting magnitude, and even when detected, such groups are typically found only at very low redshifts, where the survey volume is small and statistical uncertainties are larger. As a result, the observed HMF underestimates the true abundance of low-mass haloes, leading to an artificial drop-off in this range. Therefore, the HMF presented here should be regarded as a demonstration of the methodology rather than a definitive measurement.

To ensure reliable binning of the HMF, the provided HMFs includes groups with $N_{\mathrm{m}} \geq 3$ detected in each survey (SGP-2dF, G23-2dF, and G23-GAMA). For the SGP-2dF sample, an additional restriction to $z < 0.08$ is imposed to minimise the effects of declining completeness at higher redshifts. The HMFs shown in Figure~\ref{fig:Obs_HMF} are derived using both the standard (left panel: Equation~\ref{Eq:OG_VT}) and MVT (right panel: Equation~\ref{Eq:Cor_VT}) estimators. As discussed in Section~\ref{subsubsec:VT_VT} and illustrated in Figure~\ref{fig:Disp_HM}, the standard virial theorem estimator predicts an excess of very low-mass group haloes. When considering the observational survey volume and the presence of a local void at $z \sim 0.02$ \citep{Driver2022b, VanKempen2024}, such low-mass systems are not expected to be sampled within this dataset. This underestimation of halo masses by the virial theorem could be misinterpreted as a genuine feature in the absence of this prior knowledge. In contrast, the MVT corrects for this bias, yielding a halo mass function that is more consistent with theoretical and practical expectations, providing a more reliable representation of the underlying group halo mass distribution.

The resulting HMFs are compared both to the analytic prediction for the adopted cosmology of S\textsc{hark} and to the empirical fit derived by \cite{Driver2022}. The empirical fit from \cite{Driver2022} is based on halo masses drawn from three low-redshift group/cluster catalogues: the GAMA group catalogue of \citet{Robotham2011}, the SDSS Data group catalogue of \citet{Tempel2017}, and  ROSAT-ESO Flux Limited X-ray Galaxy Cluster Survey \citep[REFLEX II;][]{Boehringer2013, Boehringer2014, Boehringer2017}. Together these span the mass range from galaxy groups to massive clusters, enabling a continuous measurement of the $z=0$ HMF. It is important to note that the dataset used by \citet{Driver2022} to establish their empirical relation is substantially larger and homogenised compared the sample analysed in this work. Nevertheless, our non-idealised and considerably smaller dataset yields a HMF that aligns well with theoretical expectations at the high to moderate halo mass range ($\log M_{halo}>12.75~[M_{\odot}]$) and underscores the significant progress made in the development of accurate halo mass estimators. However, at lower halo masses, the observed HMFs exhibit clear signs of incompleteness, further emphasising the necessity for larger and more homogeneous samples to robustly characterise this regime and improve observational constraints on the low-mass end.

\begin{figure*}[t]
\centering
\includegraphics[width=\textwidth]{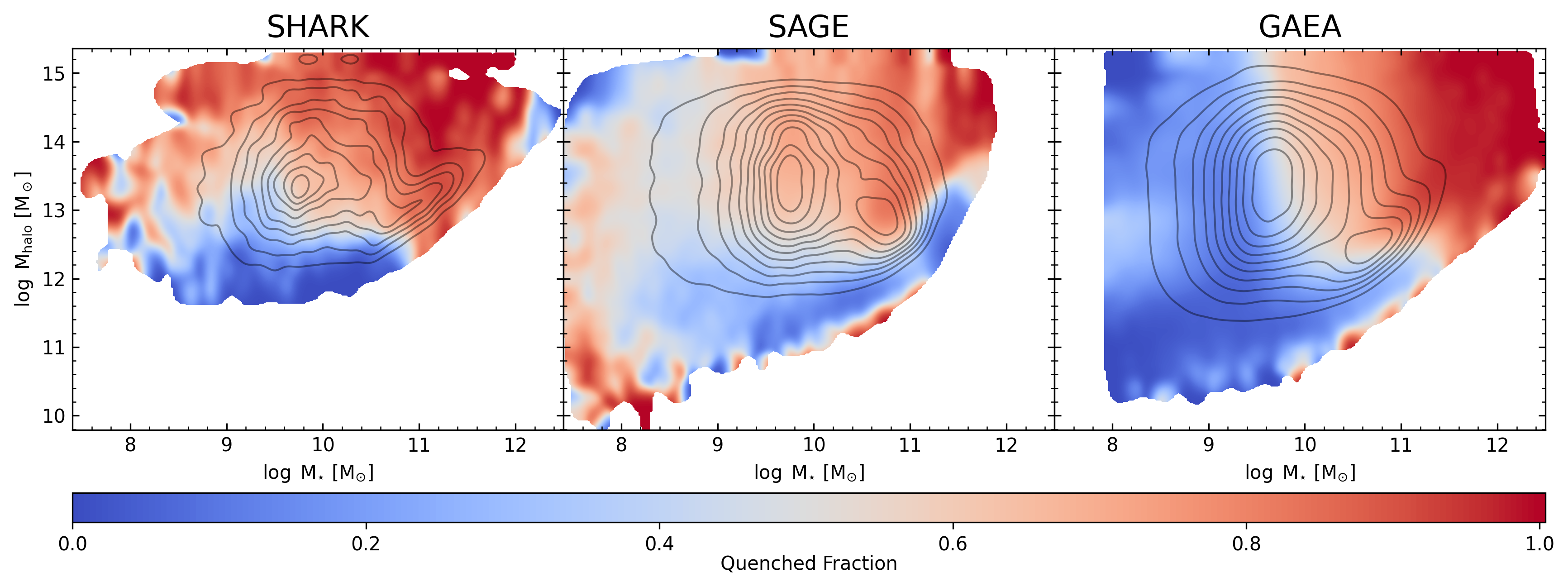}
\caption{Distribution of quenched fraction in the stellar mass–halo mass plane for galaxy groups in the SHARK, SAGE, and GAEA simulations. The colour scale indicates the fraction of quenched galaxies (defined as those with $\log\,\mathrm{sSFR} < -11.0$) within each region of parameter space, ranging from blue (predominantly star-forming) to red (predominantly quenched). The underlying distribution is computed using a two-dimensional kernel density estimate (KDE) with adaptive smoothing. The quenched fraction is calculated as the ratio of the KDE-weighted density of quenched galaxies to the total KDE-weighted density in each cell. Black contours show the underlying density of individual galaxies. Halo masses are estimated using the calibrated relation between the sum of the stellar masses of the three most massive group galaxies and the group halo mass.}
\label{fig:Sim_QF}
\end{figure*}

Looking ahead, upcoming wide-area spectroscopic surveys such as WAVES \citep{Driver2019} and 4HS \citep{ENTaylor23} will provide the depth, completeness, and sky coverage required to construct high-precision HMFs over a wide mass range. With sufficiently large samples and accurate halo mass estimates, such surveys will enable the use of the HMF as a cosmological probe, constraining parameters such as $\Omega_m$, $\sigma_8$, and potentially the nature of dark energy \citep[e.g.,][]{Murray2013, Bocquet2019}. The principal limiting factor for future HMF-based cosmological analyses is likely to be the uncertainty in group identification and membership assignment, particularly at low multiplicity, rather than the precision of the halo mass estimator itself.

%%%%%%%%%%%%%%%%%%%%%%%%%%%%%%%%%%%%%%%%%%%%%%%%%%%%%%%%%%%%%%%%%%%%%%%%%%%%%%%%%%%%%%%%%%
\subsection{Quenched Fraction in the Stellar Mass–Halo Mass Plane}

\begin{figure}[!hbt]
\centering
\includegraphics[width=\textwidth]{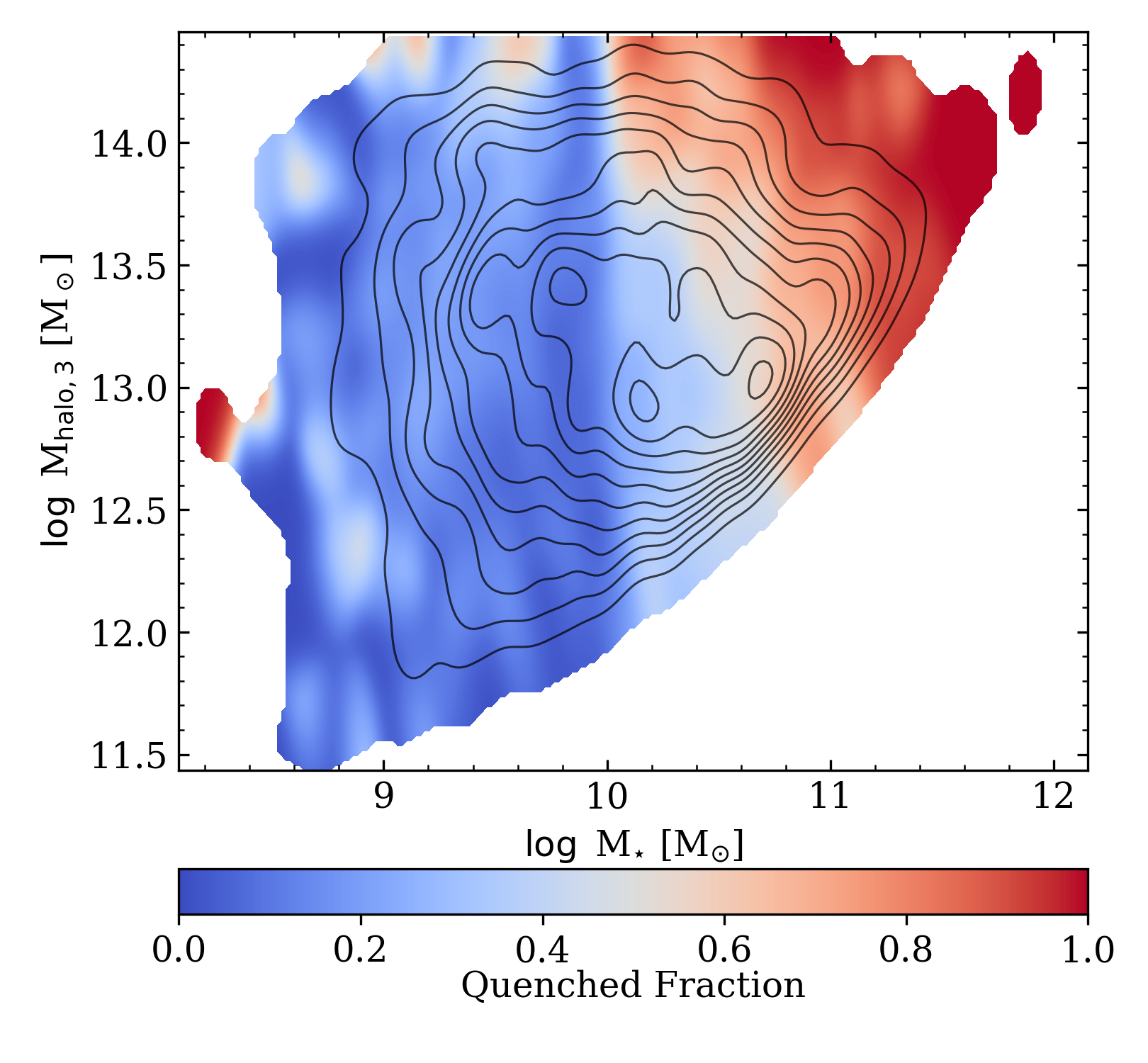}
\caption{
Distribution of quenched fraction in the stellar mass–halo mass plane for galaxy groups in the observational dataset of \citet{VanKempen2024}. The colour scale represents the fraction of quenched galaxies (defined as those with \(\log\,\mathrm{sSFR} < -11.0\)) within each region of parameter space, ranging from blue (predominantly star-forming) to red (predominantly quenched). The underlying distribution is computed using a two-dimensional kernel density estimate (KDE) with adaptive smoothing. The quenched fraction is calculated as the ratio of the KDE-weighted density of quenched galaxies to the total KDE-weighted density in each cell. Black contours indicate the underlying density of individual galaxies. The halo masses used in this figure are estimated using the calibrated sSHMR between the sum of the stellar masses of the three most massive group galaxies and the group halo mass. This figure highlights the dependence of quenching on both stellar and halo mass, with the quenched fraction increasing towards higher masses in both dimensions.
}
\label{fig:QF}
\end{figure}

The second application demonstrates the utility of the sSHMR estimator for investigating the interplay between galaxy quenching, stellar mass, and environment. To ensure compatibility with the sSHMR calibration detailed in Section~\ref{subsubsec:SHMR_Acc_Unc}, we applied a systematic offset of +0.13 dex to all observational stellar mass estimates, bringing them into alignment with the ProSpect like stellar masses used in the S\textsc{hark} calibration. This adjustment is essential for accurate halo mass estimation using Equation~\ref{Eq:SHMR} (see Section~\ref{subsubsec:SHMR_Acc_Unc} for more details). Figure~\ref{fig:Sim_QF} presents the distribution of the quenched fraction in the stellar mass–halo mass plane for the mock datasets, while Figure~\ref{fig:QF} shows the corresponding distributions for the observational dataset. The quenched fraction is defined as the fraction of galaxies with $\log\,\mathrm{sSFR} < -11.0~[\mathrm{yr}^{-1}]$, following the criteria established in \citet{VanKempen2024}, with observational SFR upper limits classified as quenched if they meet the W3 detection limit for their given distance, otherwise conservatively classified as star-forming (see \citet{VanKempen2024} for full details regarding upper-limit handling).

The underlying distribution is computed using a two-dimensional kernel density estimate (KDE) with adaptive smoothing. The KDE is evaluated on a regular grid with the kernel full width at half maximum chosen to be approximately twice the typical uncertainty in the stellar and halo mass ($0.15$ dex and $0.3$ dex, respectively). This approach reduces bias from mass errors (stellar and halo) in the smoothed contours, ensuring that the observed trends are not artificially sharpened by noise or underestimated uncertainties. The quenched fraction in each cell is calculated as the ratio of the KDE-weighted density of quenched galaxies to the total KDE-weighted density, providing a robust estimate even in regions of low sampling density. Black contours indicate the underlying density of individual galaxies for reference.

The simulated quenched fraction distributions in Figure~\ref{fig:Sim_QF} highlight the diversity in how different galaxy formation models implement satellite quenching. Notably, the GAEA simulation shows a striking dependence of the quenched fraction almost exclusively on stellar mass, with minimal variation as a function of halo mass, suggesting that internal processes dominate quenching in this model. In contrast, both SHARK and SAGE display more complex dependencies, with varying degrees of environmental influence on quenching. In the observational data, there is a pronounced transition in the quenched fraction at $M_\star \sim 10^{10}~M_\odot$, indicating a strong stellar mass threshold for quenching, alongside a secondary, smaller dependence on halo mass. These comparisons underscore the need for improved observational constraints to better calibrate and distinguish between quenching models, particularly in the treatment of environmental effects.

Looking ahead, the larger and more complete samples provided by upcoming surveys such as WAVES and 4HS will enable this analysis to be extended in greater detail, including the incorporation of cosmic web location and other environmental metrics. This will provide new opportunities to disentangle the relative roles of internal and external quenching mechanisms in galaxy evolution.

In summary, these examples illustrate the power and flexibility of the calibrated halo mass estimators developed in this work for a range of observational applications, from cosmological tests with the HMF to detailed studies of galaxy evolution in group environments.

\section{Summary and Conclusions}
\label{sec:conclusion}

Accurate estimation of dark-matter halo masses for galaxy groups is fundamental both for studies of galaxy evolution and for exploiting group catalogues as cosmological probes. In this work we developed, calibrated, and validated two complementary, observational halo-mass estimators: a modified virial-theorem estimator (MVT) and a summed stellar–halo mass relation (sSHMR) that uses the sum of the stellar masses of the three most massive group members, as the predictor. The estimators were constructed using realistic mock light cones and survey selections (Sections~\ref{subsec:Meth1} \& \ref{subsec:Meth2}); both were calibrated on the fiducial S\textsc{hark} SAM and subsequently validated on SAGE and GAEA catalogues to quantify model dependence. Performance was quantified in terms of accuracy, uncertainty, and variations between models (Sections \ref{subsubsec:VT_ACC_Unc}, \ref{subsubsec:SHMR_Acc_Unc}); the MVT was designed to minimise model dependence while the sSHMR provides the lowest scatter where reliable stellar masses are available. Finally, the practical utility of the calibrated estimators was demonstrated in two observational applications, the empirical halo mass function and the mapping of quenched fractions in the stellar mass–halo mass plane (Section~\ref{sec:ObsApp}).

The primary results of this study are as follows:
\begin{enumerate}
    \item \textbf{Modified Virial Theorem:} We present a robust, physically motivated modification of the virial theorem for group-scale halo mass estimation, incorporating corrections for velocity dispersion and projected radius (Section~\ref{subsec:Meth1}). The calibrated estimator achieves negligible systematic bias (mean $\Delta \sim 0.01$ dex) and moderate scatter ($\sigma \sim 0.23$--$0.25$ dex) across all tested SAMs, with minimal dependence on baryonic physics (Section~\ref{subsubsec:VT_ACC_Unc}, Figure~\ref{fig:Disp_HM}).
    
    \item \textbf{summed Stellar-to-Halo Mass Relation:} We calibrate an empirical relation between the sum of the stellar masses of the three most massive group galaxies and the halo mass, optimised via Bayesian MCMC (Section~\ref{subsec:Meth2}). This method yields the highest precision among those tested, with typical scatter as low as $0.12$ dex in the primary calibration sample, but exhibits greater model dependence due to its sensitivity to the underlying baryonic physics and stellar mass assembly (Section~\ref{subsubsec:SHMR_Acc_Unc}, Figure~\ref{fig:SHMR}).
    
    \item \textbf{Comparison Across Models:} Both estimators were callibrated to S\textsc{hark} and cross-validated with SAGE, and GAEA, demonstrating robust performance. The virial theorem calibration is largely insensitive to the details of the SAM, while the sSHMR shows greater variance, particularly in models with differing feedback and mass resolution (notably GAEA at low masses; Section~\ref{subsubsec:VT_ACC_Unc}, Section~\ref{subsubsec:SHMR_Acc_Unc}).
    
    \item \textbf{Observational Applications:} We showcase two key applications: (i) the construction of the empirical HMF using the calibrated virial theorem, demonstrating the feasibility of this approach for future spectroscopic surveys (Section~\ref{sec:ObsApp}, Figure~\ref{fig:Obs_HMF}); and (ii) the mapping of the quenched fraction in the stellar mass–halo mass plane using sSHMR-based halo masses, revealing the joint dependence of quenching on both stellar and halo mass (Section~\ref{sec:ObsApp}, Figure~\ref{fig:QF}).
    
    \item \textbf{Guidance for Future Work:} For cosmological applications, such as HMF construction and parameter inference, the calibrated virial theorem is recommended due to its minimal model dependence. For studies focused on group-scale halo mass estimation and environmental effects, the sSHMR provides the highest precision, provided reliable stellar mass measurements are available (Section~\ref{subsubsec:VT_ACC_Unc}, Section~\ref{subsubsec:SHMR_Acc_Unc}, Section~\ref{sec:ObsApp}).
\end{enumerate}

In conclusion, the calibrated halo mass estimators developed in this work provide a robust foundation for future analyses of group catalogues in both simulated and observational contexts. Their application will be particularly valuable for upcoming wide-area spectroscopic surveys, enabling precise studies of galaxy evolution and the dark matter halo population, and facilitating the use of group catalogues as cosmological probes.

\section{Acknowledgements}
\label{sec:Acknowledgements}

We thank the anonymous referee for helpful comments and suggestions that have improved the content and clarity of this paper. M.E.C. is a recipient of an Australian Research Council Future Fellowship (project No. FT170100273) funded by the Australian Government. D.J.C. is a recipient of an Australian Research Council Future Fellowship (project No. FT220100841) funded by the Australian Government. This publication makes use of data products from the Wide-field Infrared Survey Explorer, which is a joint project of the University of California, Los Angeles, and the Jet Propulsion Laboratory/California Institute of Technology, funded by the National Aeronautics and Space Administration. GAMA is a joint European-Australasian project based around a spectroscopic campaign using the Anglo-Australian Telescope. The GAMA input catalogue is based on data taken from the Sloan Digital Sky Survey and the UKIRT Infrared Deep Sky Survey. Complementary imaging of the GAMA regions is being obtained by a number of independent survey programmes including GALEX MIS, VST KiDS, VISTA VIKING, WISE, Herschel-ATLAS, GMRT and ASKAP providing UV to radio coverage. GAMA is funded by the STFC (UK), the ARC (Australia), the AAO, and the participating institutions. The GAMA website is \url{https://www.gama-survey.org/}. Based on observations made with ESO Telescopes at the La Silla Paranal Observatory under programme ID 177.A-3016. This research has made use of {\tt python} (\url{https://www.python.org}) and python packages: {\tt astropy} \citep{Astropy13,Astropy18, Astropy2022}, {\tt cmasher} \citep{cmasher}, {\tt emcee} \citep{emcee}, {\tt matplotlib} \url{http://matplotlib.org/} \citep{Hunter07}, {\tt NumPy} \url{http://www.numpy.org/} \citep{Walt11}, {\tt Pandas} \citep{Pandas}, and {\tt SciPy} \url{https://www.scipy.org/} \citep{Virtanen20}.

\bibliography{ref}

\appendix

\section{Virial Theorem Fitting Parameters}
\label{sec:VT Fitting}

\renewcommand{\thetable}{A\arabic{table}}
\renewcommand{\thefigure}{A\arabic{figure}}

\begin{figure}[!hbt]
\centering
\includegraphics[width=\linewidth]{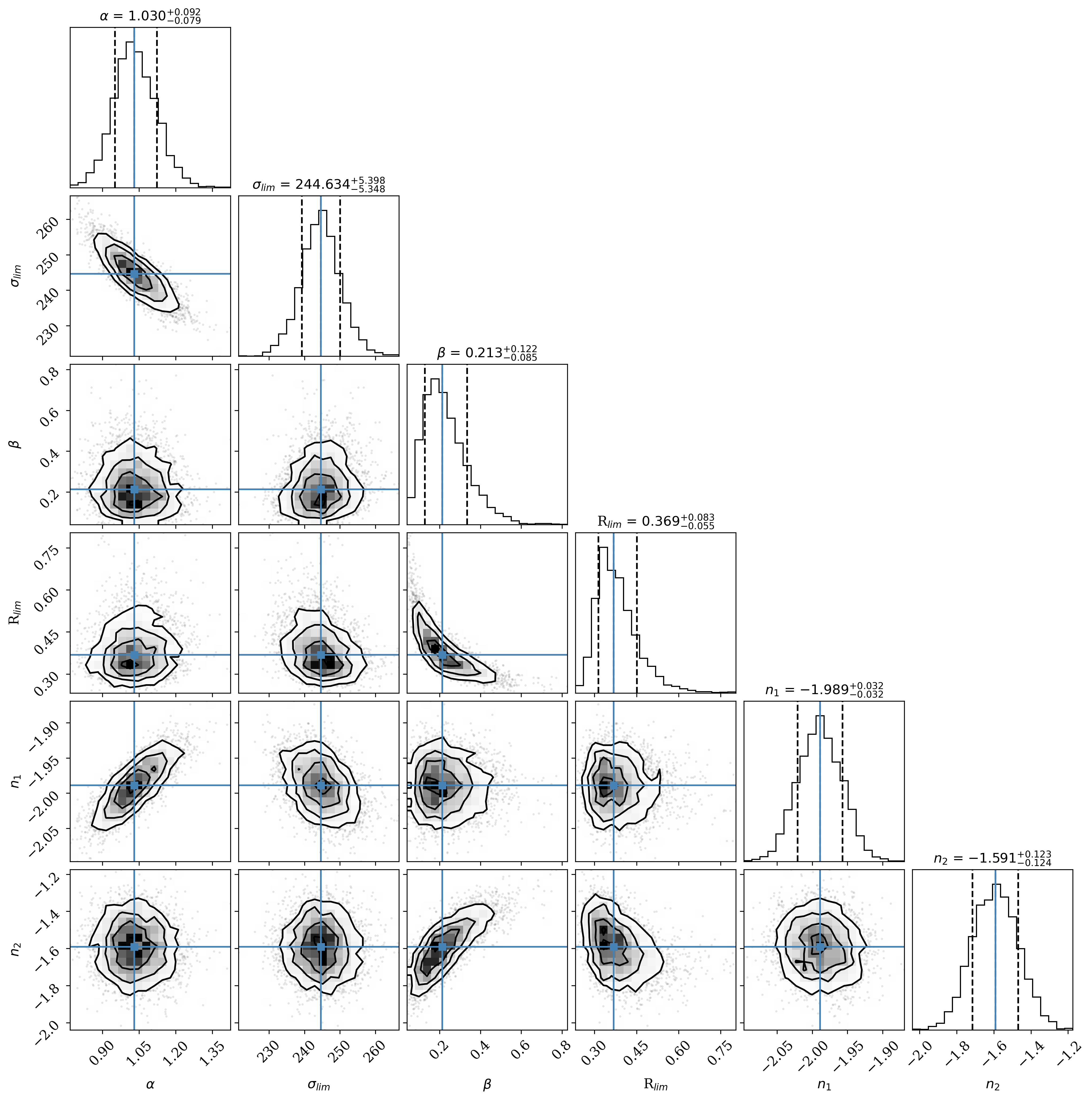}
\caption{
Posterior distributions and covariances for the six free parameters of the calibrated dispersion-based halo mass relation, as determined by MCMC sampling on the S\textsc{hark} model. The corner plot displays the marginalised one-dimensional distributions along the diagonal, with median values and 68\% credible intervals indicated, and the two-dimensional projections of the posterior for each parameter pair in the off-diagonal panels. The parameters correspond to the normalisation and scaling of the velocity dispersion and maximum projected separation terms ($\alpha$, $\sigma_{\mathrm{lim}}$, $\beta$, $R_{\mathrm{lim}}$, $n_1$, $n_2$) in the power-law model for halo mass. The MCMC analysis was performed using the \texttt{emcee} sampler, with convergence assessed via autocorrelation time. This figure demonstrates that all parameters are well-constrained and highlights the correlations between them, providing a robust statistical foundation for the calibrated group halo mass estimator.
}
\label{fig:MCMC_Disp}
\end{figure}

The calibration of the MVT was performed using a Bayesian MCMC approach, enabling robust exploration of the posterior probability distributions for all model parameters. The likelihood function was constructed to minimise the scatter in $\log M_{\mathrm{halo}}$ at fixed predicted mass, and is defined as
\begin{equation}
    \ln \mathcal{L} = -\frac{1}{2} \sum_{i} \left[ \frac{(y_i - \mu_i)^2}{\sigma^2} + \ln(2\pi \sigma^2) \right],
\end{equation}
where $y_i$ is the true halo mass, $\mu_i$ is the predicted halo mass from the calibrated virial theorem, and $\sigma$ is the standard deviation of the residuals. The model incorporates a power-law correction to the virial coefficient, parameterised as described in Section~\ref{subsec:Meth1}, with the functional form:
\begin{equation}
    A = \frac{5}{3} + \alpha \left[ \left( \frac{\sigma}{\sigma_{\mathrm{lim}}} \right)^{n_1} - 1 \right] + \beta \left[ \left( \frac{R}{R_{\mathrm{lim}}} \right)^{n_2} - 1 \right],
\end{equation}
where $\sigma$ is the group velocity dispersion, $R$ is the projected group radius, and $\alpha$, $\sigma_{\mathrm{lim}}$, $\beta$, $R_{\mathrm{lim}}$, $n_1$, and $n_2$ are free parameters.

Uniform priors were adopted for all parameters within physically motivated bounds. The MCMC sampling was performed using an ensemble sampler with 40 walkers and up to $2 \times 10^6$ steps, with convergence assessed via the integrated autocorrelation time. Burn-in and thinning were determined dynamically based on the final autocorrelation time, ensuring that the posterior samples are independent and representative of the converged distribution.

The resulting posterior distributions and covariances for the six free parameters are shown in Figure~\ref{fig:MCMC_Disp}. The median values and 68\% confidence intervals for each of the coefficients are are shown in Table~\ref{Tab:Disp_Coeff_full}.

\begin{table}[!htb]
\caption{Best-fitting of the coefficients for the modified virial theorem relation, including their 16th and 84th percentile scatter.}
\centering
\begin{tabularx}{\columnwidth}{ CCCC }
\hline \hline 
Coefficient  &  $16^{th}$ Percentile & Median &  $84^{th}$ Percentile \\ 
\hline
$\alpha$  & 0.951 & 1.030 & 1.122\\
$\sigma_{lim}[\mathrm{km/s}]$ & 239.286 &  244.634 & 250.032\\
$n_{1}$ & -2.021 & -1.989 & -1.957\\
$\beta$ & 0.128 & 0.213 & 0.335\\
$R_{lim}[\mathrm{Mpc}]$ & 0.314 & 0.369 & 0.452\\
$n_{2}$ & -1.715 & -1.591 & -1.468\\
\hline \hline
\end{tabularx}
\label{Tab:Disp_Coeff_full}
\end{table}

The corner plot in Figure~\ref{fig:MCMC_Disp} illustrates that all parameters are well-constrained, with physically plausible correlations and no significant degeneracies. This calibration provides a statistically robust foundation for the application of the virial theorem-based halo mass estimator to both simulated and observational group catalogues. For further details on the calibration methodology and performance, see Sections~\ref{subsec:Meth1} and~\ref{subsubsec:VT_ACC_Unc}.

\section{SHMR Fitting Parameters}
\label{sec:SHMR Fitting}

\renewcommand{\thetable}{B\arabic{table}}
\renewcommand{\thefigure}{B\arabic{figure}}

\begin{figure}[!hbt]
\centering
\includegraphics[width=\linewidth]{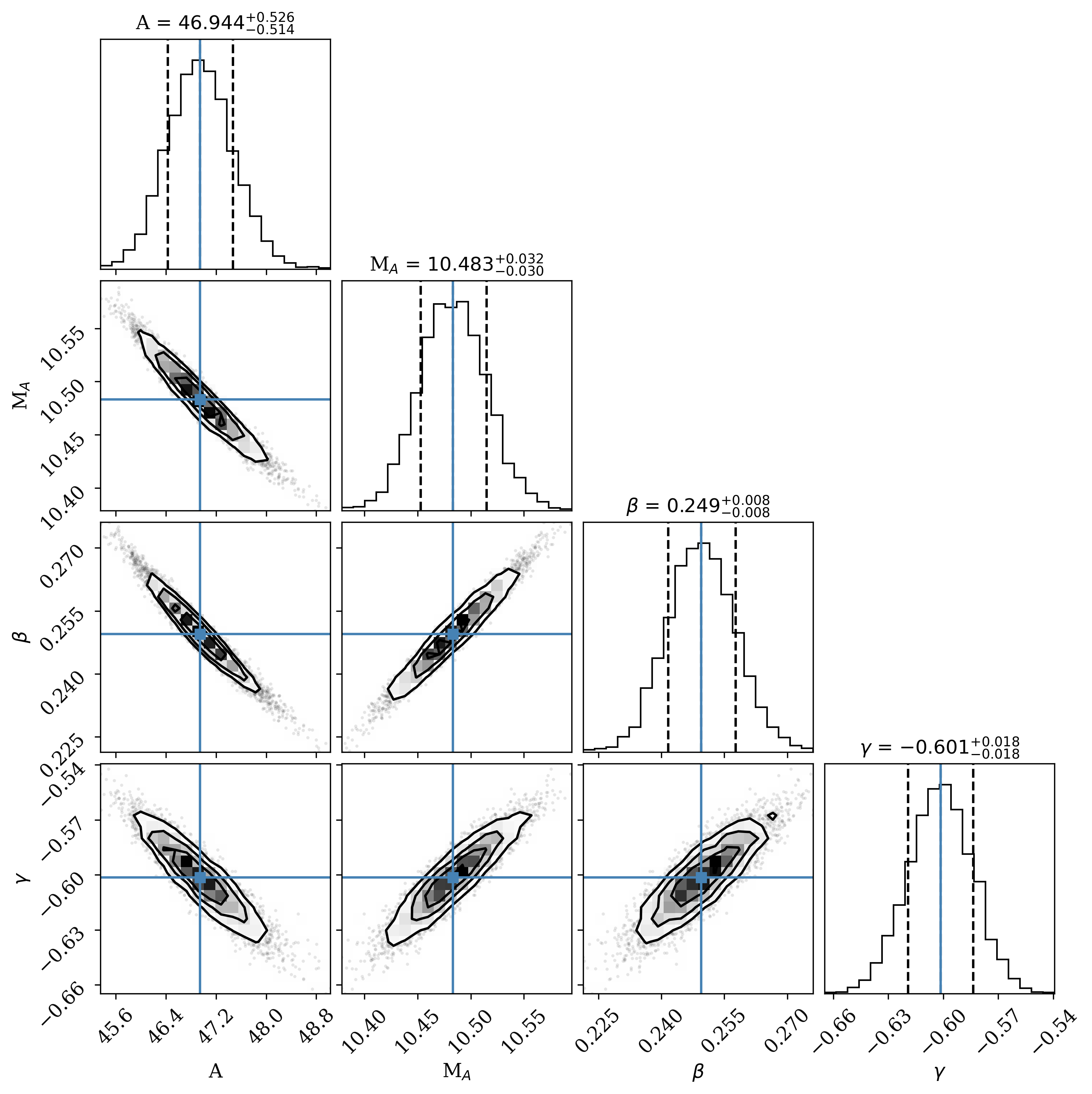}
\caption{
Posterior distributions and covariances for the four free parameters of the summed stellar mass–halo mass relation, as determined by MCMC sampling on the S\textsc{hark} model. The corner plot displays the marginalised one-dimensional posterior distributions for each parameter (normalisation $A$, characteristic mass $M_A$, low-mass slope $\beta$, and high-mass slope $\gamma$) along the diagonal, with median values and 68\% credible intervals indicated. Off-diagonal panels show the two-dimensional projections of the posterior for each parameter pair, highlighting correlations and degeneracies. The MCMC analysis was performed using the \texttt{emcee} sampler, with convergence assessed via autocorrelation time. This figure demonstrates that all parameters are well-constrained, providing a robust statistical foundation for the calibrated relation between the sum of the stellar masses of the three most massive group galaxies and their host halo mass.
}
\label{fig:MCMC_P3M}
\end{figure}

The calibration of the sSHMR was performed using a Bayesian MCMC approach, enabling a rigorous exploration of the posterior probability distributions for all model parameters. The adopted functional form for the SHMR is a double power-law,
\begin{equation}
    M_{\mathrm{halo}} = A\, \Sigma M_{*,\mathrm{3}} \left[ \left( \frac{\Sigma M_{*,\mathrm{3}}}{{M_A}} \right)^{\beta} + \left( \frac{\Sigma M_{*,\mathrm{3}}}{{M_A}} \right)^{\gamma} \right],
\end{equation}
where $M_{*,\mathrm{sum}}$ is the sum of the stellar masses of the three most massive galaxies in the group, $A$ is the normalisation, $M_A$ is the characteristic stellar mass, and $\beta$ and $\gamma$ are the low- and high-mass slopes, respectively.

The likelihood function was constructed to minimise the scatter in $\log M_{\mathrm{halo}}$ at fixed $\Sigma M_{\star, \mathrm{3}}$, and is given by
\begin{equation}
    \ln \mathcal{L} = -\frac{1}{2} \sum_{i} \left[ \frac{(y_i - \mu_i)^2}{\sigma^2} + \ln(2\pi \sigma^2) \right],
\end{equation}
where $y_i$ is the true halo mass, $\mu_i$ is the predicted halo mass from the SHMR, and $\sigma$ is the standard deviation of the residuals. Uniform priors were adopted for all parameters within physically motivated bounds.

MCMC sampling was performed using an ensemble sampler with 40 walkers and up to $2 \times 10^6$ steps, with convergence assessed via the integrated autocorrelation time. Burn-in and thinning were determined dynamically based on the final autocorrelation time, ensuring that the posterior samples are independent and representative of the converged distribution.

The resulting posterior distributions and covariances for the four free parameters are shown in Figure~\ref{fig:MCMC_P3M}. The median and the $16^{th}$ and $84^{th}$ percentile for each parameter given in Table~\ref{Tab:sSHMR_Coeff_full}.

\begin{table}[!htb]
\caption{Best-fitting of the coefficients for the sSHMR, including their 16th and 84th percentile.}
\centering
\begin{tabularx}{\columnwidth}{ CCCC }
\hline \hline 
Coefficient  &  $16^{th}$ Percentile & Median &  $84^{th}$ Percentile \\ 
\hline
$A$  & 46.43 & 46.944 & 47.47\\
$\log M_A[M_{\odot}]$ & 10.453 &  10.483 & 10.515\\
$\beta$ & 0.241 & 0.249 & 0.257\\
$\gamma$ & -0.573 & -0.555 & -0.537\\
\hline \hline
\end{tabularx}
\label{Tab:sSHMR_Coeff_full}
\end{table}

The corner plot in Figure~\ref{fig:MCMC_P3M} demonstrates that all parameters are well-constrained, with physically plausible correlations and no significant degeneracies. This calibration provides a statistically robust foundation for the application of the SHMR-based halo mass estimator to both simulated and observational group catalogues. For further details on the calibration methodology and performance, see Sections~\ref{subsec:Meth2} and~\ref{subsubsec:SHMR_Acc_Unc}.

\end{document}